\documentclass[
  journal=largetwo,
  manuscript=article-type,
  year=2020,
  volume=37,
]{cup-journal}
\usepackage{ulem}
\usepackage{float}
\usepackage{amsmath}
\usepackage[nopatch]{microtype}
\usepackage{booktabs}
\usepackage[colorlinks=true,linkcolor=blue, citecolor=blue, allcolors=blue]{hyperref}

\newcommand{\HI}{\rm H{\scriptsize I}~}

\title{Measuring the global 21-cm signal with the MWA-II: improved characterisation of lunar-reflected radio frequency interference}

\author{Himanshu Tiwari}
\affiliation{International Centre for Radio Astronomy Research (ICRAR), Curtin University, Kent Street, Bentley, Perth, Western Australia, 6102.}
\alsoaffiliation{Commonwealth Scientific and Industrial Research Organisation (CSIRO), Space \& Astronomy, P. O. Box 1130, Bentley, WA 6102, Australia.}
\email[Himanshu Tiwari]{himanshu.tiwari@postgrad.curtin.edu.au}

\author{Benjamin McKinley}
\affiliation{International Centre for Radio Astronomy Research (ICRAR), Curtin University, Kent Street, Bentley, Perth, Western Australia, 6102.}
\alsoaffiliation{ARC Centre of Excellence for All Sky Astrophysics in Three Dimensions (ASTRO-3D), Australia.}

\author{Cathryn M. Trott}
\affiliation{International Centre for Radio Astronomy Research (ICRAR), Curtin University, Kent Street, Bentley, Perth, Western Australia, 6102.}
\alsoaffiliation{ARC Centre of Excellence for All Sky Astrophysics in Three Dimensions (ASTRO-3D), Australia.}

\author{Nithyanandan Thyagarajan}
\affiliation{Commonwealth Scientific and Industrial Research Organisation (CSIRO), Space \& Astronomy, P. O. Box 1130, Bentley, WA 6102, Australia.}

\keywords{Early Universe, Cosmic Dawn and Reionization, first stars, observations, MWA} 

\begin{document}

\begin{abstract}
    Radio interferometers can potentially detect the sky-averaged signal from the Cosmic Dawn (CD) and the Epoch of Reionisation (EoR) by studying the Moon as a thermal block to the foreground sky. The first step is to mitigate the Earth-based RFI reflections (Earthshine) from the Moon, which significantly contaminate the FM band $\approx 88-110$ MHz, crucial to CD-EoR science. We analysed MWA phase-I data from $72-180$ MHz at $40$ kHz resolution to understand the nature of Earthshine over three observing nights. We took two approaches to correct the Earthshine component from the Moon. In the first method, we mitigated the Earthshine using the flux density of the two components from the data, while in the second method, we used simulated flux density based on an FM catalogue to mitigate the Earthshine. Using these methods, we were able to recover the expected Galactic foreground temperature of the patch of sky obscured by the Moon. We performed a joint analysis of the Galactic foregrounds and the Moon's intrinsic temperature $(T_{\rm Moon})$ while assuming that the Moon has a constant thermal temperature throughout three epochs. We found $T_{\rm Moon}$ to be at $184.4\pm{2.6}\rm ~K$ and $173.8\pm{2.5}\rm ~K$ using the first and the second methods, respectively, and the best-fit values of the Galactic spectral index $(\alpha)$ to be within the $5\%$ uncertainty level when compared with the global sky models. Compared with our previous work, these results improved constraints on the Galactic spectral index and the Moon's intrinsic temperature. We also simulated the Earthshine at MWA between November and December 2023 to find suitable observing times less affected by the Earthshine. Such observing windows act as Earthshine avoidance and can be used to perform future global CD-EoR experiments using the Moon with the MWA.
\end{abstract}

\section{Introduction} \label{section: Introduction}

The Cosmic Dawn (CD) marked the end of the Dark Ages with the emergence of the first luminous objects in the early Universe. These objects were the first generation of stars, black holes and other compact objects. The X-ray and UV radiation produced by these objects started heating and ionising their surrounding matter. This ionising process gradually changed the state of the IGM from neutral to fully ionised during the Epoch of Reionisation (EoR) until the redshift $z \sim 5.3$ \citep[see][for review]{Furlanetto_2004, Bharadwaj_2005, Pritchard_2012, Mesinger_2016}.
The observational evidence from high redshift quasar spectra \citep{fan03, barnett17}, Ly-$\alpha$ emitters \citep{mcquinn06} and Cosmic Microwave Background (CMB) scattering from the ionised IGM at low redshifts implies that the cosmic reionisation lasted between redshifts $z\approx 18-5.3$ \citep{komatsu11, Planck_2018_cosmo}. However, these studies have only placed weak constraints on the astrophysical properties of the first stars, black holes and galaxies and the evolution of the early Universe due to the difficulty of direct observations. The neutral medium of atomic Hydrogen (\HI) is largely opaque to UV radiation, whereas the X-ray tends to heat the IGM; therefore, it becomes challenging to observe the CD-EoR directly at these frequencies due to absorption by the IGM. An indirect approach of using the 21-cm emission line from the leftover \HI from this era is the most likely candidate to probe the CD-EoR. \citep{Mesinger_2016}

The 21-cm signal arises from the spontaneous transition of Hydrogen from the ground state triplet to the singlet state, causing the emission of a radio photon of $\approx1420$ MHz, which can be detectable by ground-based radio antennae. Therefore, several radio instruments, e.g. Murchison Widefield Array (MWA) \citep{Tingay_13}, Precision Array for Probing the Epoch of Reionization (PAPER) \citep{Pober_2011}, Hydrogen Epoch of Reionization Array (HERA) \citep{deboer17}, LOw-Frequency ARray (LOFAR) \citep{Haarlem_2013}, New Extension in Nancay upgrading LOFAR (NenuFAR) \citep{Mertens_2021}, Giant Meter wave Radio Telescope (GMRT) \citep{Paciga13}, Long Wavelength Array (LWA) \citep{Eastwood_2019} etc., aim to detect the 21-cm signal from the CD-EoR. These instruments target either the statistical features of the 21-cm signal (e.g. power spectrum) using radio interferometers, e.g. MWA \citep{trott20}, GMRT \citep{GMRT_upper_limits}, LOFAR \citep{Mertens_2020}, HERA \citep{HERA_2021}, PAPER \citep{Parsons_2014} or the volume averaged 21-cm signal (global 21-cm signal) using a single antenna, e.g. Experiment to Detect the Global EoR Signature (EDGES) \citep{Bowman_2008, Bowman_2010, Bowman_18}, Broad-band Instrument for Global HydrOgen ReioNization  Signal (BIGHORNS) \citep{Sokolowski_2015}, Shaped Antenna measurement of the background RAdio Spectrum (SARAS) \citep{patra_2012}, SARAS2 \citep{singh_2018}, SARAS3 \citep{saras3_singh2022}, Large Aperture Experiment to Detect the Dark Age (LEDA) \citep{Bernardi_2016}, Dark Ages Radio Explorer (DARE) \citep{Burns_2012} now known as Dark Ages Polarimeter Pathfinder (DAPPER) \citep{Burns_2019},  Sonda Cosmol\'ogica de las Islas para la Detecci\'on de
Hidr\'ogeno Neutro (SCI-HI) \citep{Voytek_2014}, Probing Radio Intensity at High-Z from Marion (PRIZM) \citep{prizm_17}.  However, the radiation received by these instruments is dominated by the foreground emission from the Galactic and extra-galactic radio sources. Alongside these foregrounds, instrument-based systematics, thermal noise, and radio frequency interference (RFI) present enormous challenges in 21-cm signal detection. 

The global EoR experiments measure the sky-averaged strength of the 21-cm signal by estimating the total power of 21-cm brightness as a function of frequency, which represents the redshift evolution of the early Universe. The first claimed detection of the global 21-cm signal came from EDGES \citep{Bowman_18}, which found an absorption trough profile around 78 MHz, which aligned with the theoretical predictions of early cooling and reheating \citep{Furlanetto_2006, Pritchard_2012}.
However, recent cross-validations from the SARAS3 experiment found that EDGES detection was mainly due to systematic errors \citep{saras3_singh2022}. This signifies that, despite the high signal-to-noise ratio (SNR) of these experiments, the non-thermal components of the receiver noise can mimic the cosmic signal of interest, leading to false detection scenarios. Therefore, minimisation of the instrumental systematics is the major goal of these experiments \citep{Monsalve_2017a, Singh_2017, Sokolowski_2015}.  

An alternative approach, known as the lunar occultation method, has the potential to detect the global 21-cm signal. Instead of a single antenna or dipole, this approach utilises radio interferometers to detect the sky-averaged 21-cm signal. Interferometers, in general, are not sensitive to the global sky as the response rapidly falls off for baselines  $> 1\lambda$. However, the presence of the Moon introduces a mask in the sky, which helps sustain the coherence in the global sky response at longer baselines $(> 50\lambda)$ (see fig. 1 from \citet{Vedantham_2015} for reference). The idea of lunar occultation was first proposed by \citet{Shavar_1999} and has been implemented by \citet{McKinley_2012, Ben_2018} on MWA, and \citet{Vedantham_2015} on LOFAR. In lunar occultation, the Moon is treated as a thermal block in the sky, and the interferometer measures the difference between the Moon's temperature and the global sky temperature. The benefits of using interferometers are that they contain independent antenna elements, and the voltage correlations at their receivers are devoid of frequency-dependent receiver-noise bias. Also, unlike single antenna or dipole-based experiments, which measure and mitigate the entire foreground sky to detect the global 21-cm signal, the lunar occultation method only measures the patch of sky occulted by the Moon and hence is only required to mitigate the foregrounds occulted by the Moon. This significantly reduces the Galactic emission and spectral index anomalies, which otherwise are very difficult to deal with. Therefore, lunar occultation can be used to cross-validate the findings from the EDGES and SARAS-3 experiments, and since more sensitive instruments, such as MWA phase II \& III, are currently functional, and the upcoming SKA operations are also on the horizon, it is worth further investigating the lunar occultation approach for CD-EoR science.  \par

In this work, we extended the approach of \citet{Ben_2018} by incorporating multiple nights of higher (time and frequency) resolution MWA-phase I data. We also used a new simulation-based approach to model the reflected RFI (Earthshine) in the FM band (88-110 MHz) to mitigate the reflected FM response from the Moon. Using these two approaches, we obtained improved constraints on the Moon's intrinsic temperature, the Galactic foreground temperature and spectral index. This paper is organised as follows: In \S\ref{section: Background}, we briefly summarise the lunar occultation technique. In \S\ref{section: observations} and \S\ref{section: data_processing}, we provide information about the MWA phase-I observations and data processing techniques used in this work. In \S\ref{section: Earthshine_simulation}, we describe our approach to modelling the reflected RFI from the Moon (Earthshine). In \S\ref{section: results}, we provide the main results and discuss them in \S\ref{section: discussion} along with the limitations and potential aspects of our project, and finally, we provide the conclusion in \S\ref{section: conclusion}.

\section{Background} \label{section: Background}

Lunar occultation provides a unique way to detect the EoR 21-cm global signal using radio interferometers. In the radio frequencies corresponding to the CD-EoR, the Moon is treated as a thermal source at a constant blackbody temperature. The interferometer measures the difference between the Moon's temperature and the background sky temperature of the patch of sky occulted by the Moon. This can be expressed as,
\begin{equation}\label{eq: delta_t}
    \Delta T (\nu) = T_{\rm Lunar} (\nu) - T_{\rm sky} (\nu) = \frac{c^2 S_m(\nu)}{2k_{\rm B}\nu^2\Omega} \rm~~ K
\end{equation}
where $S_m$ is the observed flux density of the Moon, $\Omega$ is the Moon's solid angle, $k_{\rm B}$ is Boltzmann's constant, $ c $ is the speed of light, and ${ \nu }$ is the frequency. The Moon reflects a part of the Galactic foreground and terrestrial radio emission back to the observer on the Earth \citep{Evans_1969}. Therefore, the Moon's temperature includes two additional factors; the reflected Galactic foreground and reflected RFI. Thus, we can express eq.\ref{eq: delta_t} as,
\begin{multline}\label{eq: temp_all}
   \Delta T (\nu) = T_{\rm Lunar} (\nu) - T_{\rm sky} (\nu)\\
    = [T_{\rm Moon} + T_{\rm refl-Earth}(\nu) + T_{\rm refl-Gal}(\nu)]\\
    - [T_{\rm Gal}(\nu) + T_{\rm CMB} + T_{\rm EoR}(\nu)],
\end{multline}
where $T_{\rm Moon}$ is the intrinsic temperature of the Moon,\\  $T_{\rm refl-Earth}(\nu)$, is the reflected RFI (Earthshine) temperature, $T_{\rm refl-Gal}(\nu)$ is the reflected Galactic temperature. The sky temperature $T_{\rm sky}(\nu)$ includes the contribution from the Galactic temperature $T_{\rm Gal}(\nu)$ (temperature of the occulted patch of the sky), Cosmic Microwave Background temperature $T_{\rm CMB}$ and EoR 21-cm global sky temperature $T_{\rm EoR}$. We used  $T_{\rm CMB}\approx 2.725 \rm~ K$ \citep{Mather_1994} in our analysis. 

In most cosmic reionisation scenarios predicted by the reionisation models, the global 21-cm signal goes through major phase transitions between 88-110 MHz; however, in this band, the Moon temperature is hugely contaminated by the Earthshine. Therefore, the first step in the detection of the global 21-cm signal would be to mitigate FM reflections from the Moon. In a similar fashion, the Galactic foregrounds would be required to be mitigated in the subsequent steps, and once all components from the Moon are successfully evaluated, the $T_{\rm EoR}$ can be isolated from the Moon.

\section{Observations} \label{section: observations}

In this work, we used six nights of MWA-phase I data from 2015. Our observational strategy was to observe the same patch of sky on two different nights at the same (Local Sidereal Time) LST, where the Moon was present on one night (ON-Moon) and absent on the other (OFF-Moon). These observations were typically separated by two to three days and observed at identical LST (LST-locked). The benefit of using these LST-locked observations is to eliminate artefacts and sidelobes from the sources in the observed field by taking the difference between the ON and OFF-Moon images. The six nights of observations were carried out on $30^{\rm th}$ August, $26^{\rm th}$ September and $21^{\rm st}$ December for the ON-Moon and on $2^{\rm nd}$ September, $29^{\rm th}$ September and $24^{\rm th}$ December $2015$ for the OFF-Moon, respectively (see observation table  \ref{tab: observation-data}). Throughout this paper, we used the nomenclature of Epochs 1,2 and 3 to represent the ON-OFF Moon (paired) datasets from August, September and December, respectively. 

The effect of ionospheric activity can cause a significant difference in the flux and positional offsets of sources over the period of two nights. However, studies of ionospheric activity have shown that the majority of these effects are minimal at the MRO \citep{Jordan_2017, Trott_2018}. We did not see a significantly bad ionospheric shift, flux anomalies or positional offsets in our data. Therefore, we ignored the effects of the ionosphere on the data. The MWA uses discrete analog beamformer settings to track the Moon and takes the observation in $30.72$ MHz contiguous bands, resetting the beamformer settings close to the Moon's location after every observation. This is known as the drift and shift tracking method \citep{Trott_2014}. A full-band $72-230$ MHz observation includes $5$ successive observations of such $30.72$ MHz contiguous bands. In this work, we used a total of $340$ observations which comprise a total of $68$ full-band observations for both ON and OFF Moon from all six nights. Since each ON-Moon observation has an LST-locked OFF-Moon pair to perform the differencing, this results in a total of $34$ full-band ON-OFF Moon observations. The individual observations were carried out for $\approx230$ seconds each.  We processed these observations using \texttt{COTTER}\footnote{\url{https://github.com/MWATelescope/cotter}} \citep{Offringa_2015} for a frequency and time resolution of $40$ kHz and $4$ seconds, respectively. The observational details are given in table \ref{tab: observation-data}.
At each observation, the central fine channel and two adjacent fine channels at the edge of each $1.28$ MHz course channel were flagged. We also flagged the combined bad dipoles/tiles from each pair of the ON-OFF Moon observations (i.e. we isolated the common working dipoles from both ON and OFF Moon observations and flagged the rest), which otherwise would have created a difference in the UV coverage on the ON and OFF-Moon observations, which could cast some additional artefacts in the final difference (ON-OFF) images.  Additionally, the operating frequencies of the ORBCOMM satellites from 121-136 MHz were avoided in our analysis.

\begin{table*}
\begin{tabular}{c|c|c|c|c|c|c|c}
    \toprule
    \multicolumn{8}{c}{$\rm ON-Moon$}\\
    \hline
     $\rm Date$  & $\rm N_{obs}$ & $\rm N_{obs}^{full-band}$ & $\rm Bandwidth~(MHz)$ & $\rm Freq.~res. ~(kHz)$ &$\rm Time ~res.~(sec)$ &  $\rm Total~ Int.~ time~ (sec)$  &  $\rm Obs.~ duration ~(hrs)$  \\
        \midrule
        $\rm 30^{th} Aug. 2015$ & $60$ &$12$&  $30.76$ &  $40$& $4$& $236$ & $3.93$  \\
        $\rm 26^{th} Sept. 2015$  & $55$ &$11$&$30.76$&  $40$& $4$ & $236$& $3.60$   \\
$\rm 21^{th} Dec. 2015$ & $55$ &$11$ &$30.76$ &  $40$& $4$& $236$  & $3.60$     \\
        \hline
        \multicolumn{8}{c}{$\rm OFF-Moon$}\\
        \hline
        $\rm 2^{nd} Sept. 2015$ & $60$ &$12$& $30.76$ &  $40$& $4$&
 $236$ & $3.93$  \\
        $\rm 29^{th} Sept. 2015$ & $55$ &$11$& $30.76$&  $40$& $4$& $236$  &  $3.60$  $\rm $ \\
        $\rm 24^{th} Dec. 2015$ & $55$ &$11$&$30.76$ &  $40$& $4$& $236$ & $3.60$  \\
        \hline
        & $\rm N_{total}=340$ & $\rm N_{total}^{full-band}=68$ & & & & & $\rm Time_{total}^{ON-Moon} =11.13$\\
    
    \bottomrule
    \end{tabular}
    \caption{ON-Moon and OFF-Moon observation details.}
    \label{tab: observation-data}
\end{table*}

\section{Data processing and modelling} \label{section: data_processing}

The processed observations were arranged according to the ON-OFF Moon pairs and phase-shifted to the precise location of the Moon to produce difference images. We used \texttt{astropy} to locate the sky position of the Moon and shifted the phase-centre of the measurements to the location of the Moon. In order to calibrate the sky visibilities, we produced the sky model using the Positional Update and Matching Algorithm (\texttt{PUMA})  catalogue \citep{line_2017} with the $800$ brightest sources in the field around the Moon. The sky model was passed to \texttt{MWA-Hyperdrive}\footnote{\url{https://github.com/MWATelescope/mwa_hyperdrive}} to perform Direction Independent (DI) calibration. The calibrated measurement sets were then used to produce primary-beam-corrected images. We used \texttt{WSCLEAN}\footnote{\url{https://gitlab.com/aroffringa/wsclean}} \citep{offringa-wsclean-2014} to produce the Stokes-I beam-corrected images from the calibrated measurement sets. Each measurement set produced $768$ images corresponding to $\approx 30.72~\rm MHz$ of bandwidth, with each image produced at $40~\rm kHz$ frequency resolution. The images were produced with $2048\times2048$ pixels, with each pixel covering approximately $0.0085^{\circ}$, and the image spanning $\approx 17^{\circ}$ of the sky on each side.  

The MWA's observational strategy uses integer delays across the MWA's $4\times4$-dipole tiles (in multiple units of $\mu$ sec.) to set the beamformers. As a result, there are a limited number of \textit{"sweet-spot"} pointing locations in the sky. Thus, in the majority of cases, the pointing centre of the beam was not at the precise sky position of the Moon during the observation. Since the primary beam size reduces with increasing frequency, most of the high-band observations were affected by the Moon being located beyond half power point of the beam (resulting in a low SNR). Therefore, we limited our analysis to the beam's full-width half maximum (FWHM), discarding observations where the Moon was located beyond the FWHM. 
As a result, we had a full-band frequency range of our processed images from $\approx 72-180$ MHz. 

\begin{figure*}[ht] 
    \centering
    \includegraphics[scale=0.45]{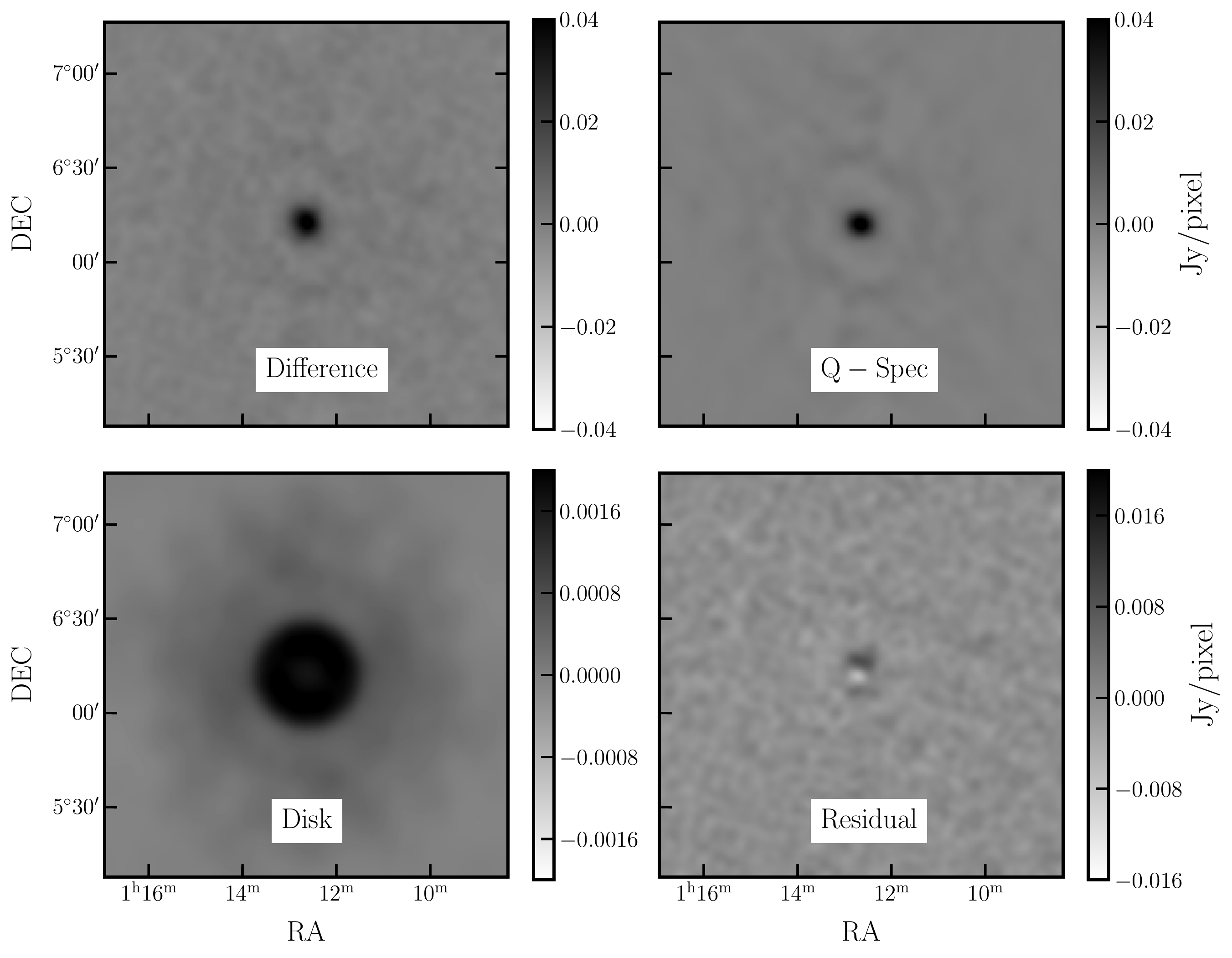}
    \caption{\{Top panel\} Left: Difference image of the ON-OFF observation, Right: The reconstructed model of the quasi-specular Earthshine component obtained by multiplying $s_{\rm spec}$ with quasi-specular mask [\textbf{B}]. \{Bottom panel\}, Left: The reconstructed disk of the Moon is obtained by operating the PSF [\textbf{P}] to the disk mask [\textbf{M}], Right: The residual image, obtained by subtracting the reconstructed disk and specular Earthshine components from the difference image. The shown images are the average of all components from the second observational Epoch (Sept. 2015) at the middle of the FM band $(\approx \nu_0=100~\rm MHz)$}.
    \label{fig: crop_image}
\end{figure*}

Finally, we took the difference between the ON and the OFF Moon images and proceeded to the flux-density estimation and Earthshine mitigation processes.\\ The detailed radar studies of the Moon by  \cite{Evans1966StudyOR, Evans_1969, Hogan_1979}.  \citep{Evans_1969} showed that the scattering of the reflected radio power can be assumed to be coming from two distinct features of the Moon's surface. The specular reflections from the smooth Moon and the diffuse reflections from the rough Moon. These studies showed that the radar cross-section area and the reflected power of the specular component decrease with increasing frequency while the increase with increasing frequency for the diffuse component.
We followed the same two-component model (previously used by \cite{Ben_2018}) to estimate the reflected RFI from the Moon. The first component was a quasi-specular Earthshine component which was a point-like source in the middle of the Moon's disk and corresponded to the reflection from the smooth surface\footnote{Moon's disk is the $2$D projection of Moon's area}, and the second component was the diffuse reflection from the rough disk of the Moon. (Note that the quasi-specular Earthshine can have a typical angular size of $\approx 16 ~\rm arcsec$ at the centre of the Moon's disk (\citet{Vedantham_2015}), whereas the diffuse is roughly equal to the Moon's angular size). The two-component Earthshine model can be described as:
\begin{equation}
    \label{eq:flux_eval}
    s_{\rm disk}, s_{\rm spec} = {\rm (H^T H)^{-1}H^T \textbf{D}}
\end{equation}
where the $s_{\rm disk}$ is the flux density (in Jy/pixels units) of the Moon's disk, $s_{\rm spec}$ is the flux density of the quasi-specular Earthshine at the centre of the Moon, and  [\textbf{D}] is the beam-corrected difference image.  The vector operator $\rm H$ is defined using the PSF [\textbf{P}] convolved over unity masks [\textbf{M}, \textbf{B}], 
\[\rm {H = \left[\textbf{M}\ast \textbf{P}~~~~ \textbf{B}\ast \textbf{P}\right]},\]
the mask \textbf{M} represents the disk, \textbf{B} represents the quasi-specular mask, \textbf{P} is the PSF of the image and $\ast$ represents the convolution operation. The disk mask has the same angular size as the Moon, whereas the quasi-specular mask occupies $8\times 8$ pixels at the centre of the Moon's image. Fig.\ref{fig: crop_image} shows the cropped $(252 \times 252)$ central pixels of the difference image, the reconstructed quasi-specular component, the disk component and the residual at the middle of the FM-band $\approx 100$ MHz ($\nu_0$, hereafter).  Finally, the estimated values of $s_{\rm disk}$ and $ s_{\rm spec}$ are summed over the corresponding masks to get the total flux density of the total disk and quasi-specular earthshine components.
\begin{equation}
    S_{\rm disk} = \sum s_{\rm disk}.{\rm \textbf{M}}; ~~S_{\rm spec} =  \sum s_{\rm spec}.{\rm \textbf{B}}
\label{eq: tot_flux}
\end{equation}
(Note that the flux density of the disk component $S_{\rm disk}$ includes the contribution of the flux density $(S_m)$of the Moon and the diffuse Earthshine $(S_{\rm diffuse})$). 
In this work, we used two different approaches to separate the Earthshine from the Moon's flux density. In the first approach, we followed a similar method as \citet{Ben_2018}, whereas in the second, we mitigated the Earthshine by simulating the FM flux density. Finally, we estimated the flux density of the Moon, $S_m(\nu)$ from $S_{\rm disk}$ as described later in \S\ref{subsection: S_m_estimate}.

\section{Earthshine Simulation}\label{section: Earthshine_simulation}

\subsection{Simulation Motivation} \label{subsection: simulation_motivation}

At radio frequencies, the Moon reflects about $7\%$ of the incident radiation falling on its surface \citep{Evans_1969}. As a result, the Moon reflects back a significant amount of the residual radiation which escapes the Earth. We can observe strong RFI reflections from the Moon in the FM band (88-110 MHz) and Digital TV (180-220 MHz). The exact behaviour of these RFI reflections is unknown, as it depends on the time and the location of the Moon and the observatory during the observation. Therefore, simulating the reflected radiation from the Moon in the context of MWA's Moon observations can be useful in understanding the nature of the reflected RFI from the Moon. In addition, the simulations can help investigate the RFI reflections at different LSTs, which can be used to identify the minimal RFI imprint during the observation window and help in scheduling future observations.

\subsection{Simulation Method} \label{subsection: simulation_method}

\begin{figure*}[ht]
    \centering
    \includegraphics[scale=0.45]{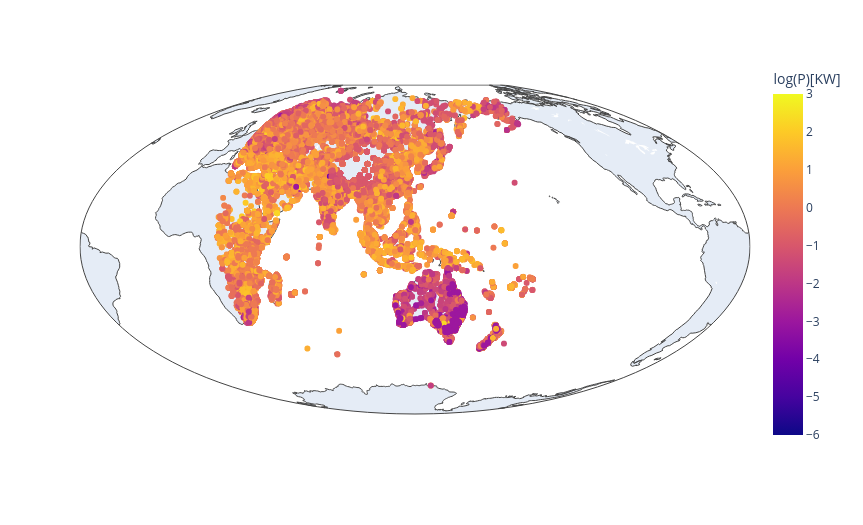}
    \caption{The locations of FM stations considered for the reflected RFI estimation during the ON-Moon observation. The figure represents a single snapshot in the middle of the second observation epoch (Sept. 2015). The colourbar represents the transmitting power of the stations in kW. The figure represents all the stations from where the Moon is above the horizon at the time of ON-Moon MWA observation.}
    \label{fig: world_FM_map}
\end{figure*}

In this analysis, we used a catalogue of FM stations\footnote{\url{https://fmlist.org}} to estimate the reflected Earthshine from the Moon. The catalogue contains information on the location of $\approx 171,000$ FM stations across the Earth, including their operating frequencies and transmitting powers. In modelling the simulated Earshshine, we made a few assumptions based on inadequate information on the beam pattern and operating hours of the FM stations in the catalogue. We assumed that all FM stations transmit isotropically throughout the full day. Due to the isotropic beam assumption, the reflected RFI includes the contribution from all of the FM stations where the Moon was above the horizon at their location at the time of ON-Moon observation. Fig.\ref{fig: world_FM_map} shows the location of all the stations appearing above the horizon as viewed from the Moon at the time of observation on the world map.

FM broadcasting follows different operating standards in different countries; therefore, the bandwidth of FM stations varies worldwide. We followed the conservative approach of Carson's rule \citep{carson} to estimate the bandwidth of FM stations and assumed that all FM stations have the same frequency deviation of $75\rm~ kHz$ with a modulation frequency of $15 \rm~ kHz$. It provided us with a bandwidth $\Delta B$ of $180~\rm kHz$ (see Chapter 4 from \citet{Haykin2007} for reference). We assumed all stations transmit a constant power across $\Delta B$. Finally, we estimated the reflected FM flux density from all the contributing FM stations at the MWA location using the following equation:
\begin{equation} \label{eq: fm_flux}
    S_{{\rm FM}(\nu)}= \frac{P_{\rm rec}(\nu)}{ \Delta B A_{\rm eff}},
\end{equation}
where $P_{\rm rec}(\nu)$ is the received power defined as radar equation (eq. 24 \cite{Evans_1969})

\begin{equation*} \label{eq: power_rec}
    P_{\rm rec}(\nu)= \frac{P_{\rm emit}(\nu)\sigma_{\rm cross}A_{\rm eff}}{{(4 \pi)}^2 D_1^2D_2^2 },
\end{equation*}
where $P_{\rm emit}(\nu)$ is the transmitted power, $D_1$ is the distance between the FM station to the Moon, $D_2$ is the distance from the Moon to the MWA, $\Delta B$ is the transmission bandwidth, and $A_{\rm eff}$ is the effective area of the MWA telescope. $\sigma_{\rm cross} = 0.081\pi R_{\rm Moon}^2$ is the radar cross-section area; it includes the contribution of $7\%$ of Moon's albedo \citep{Evans_1969}. We sampled our simulation at every $40$ kHz to match the frequency resolution of the data.
\begin{figure}[ht]
    \centering
    \includegraphics[scale=0.38]{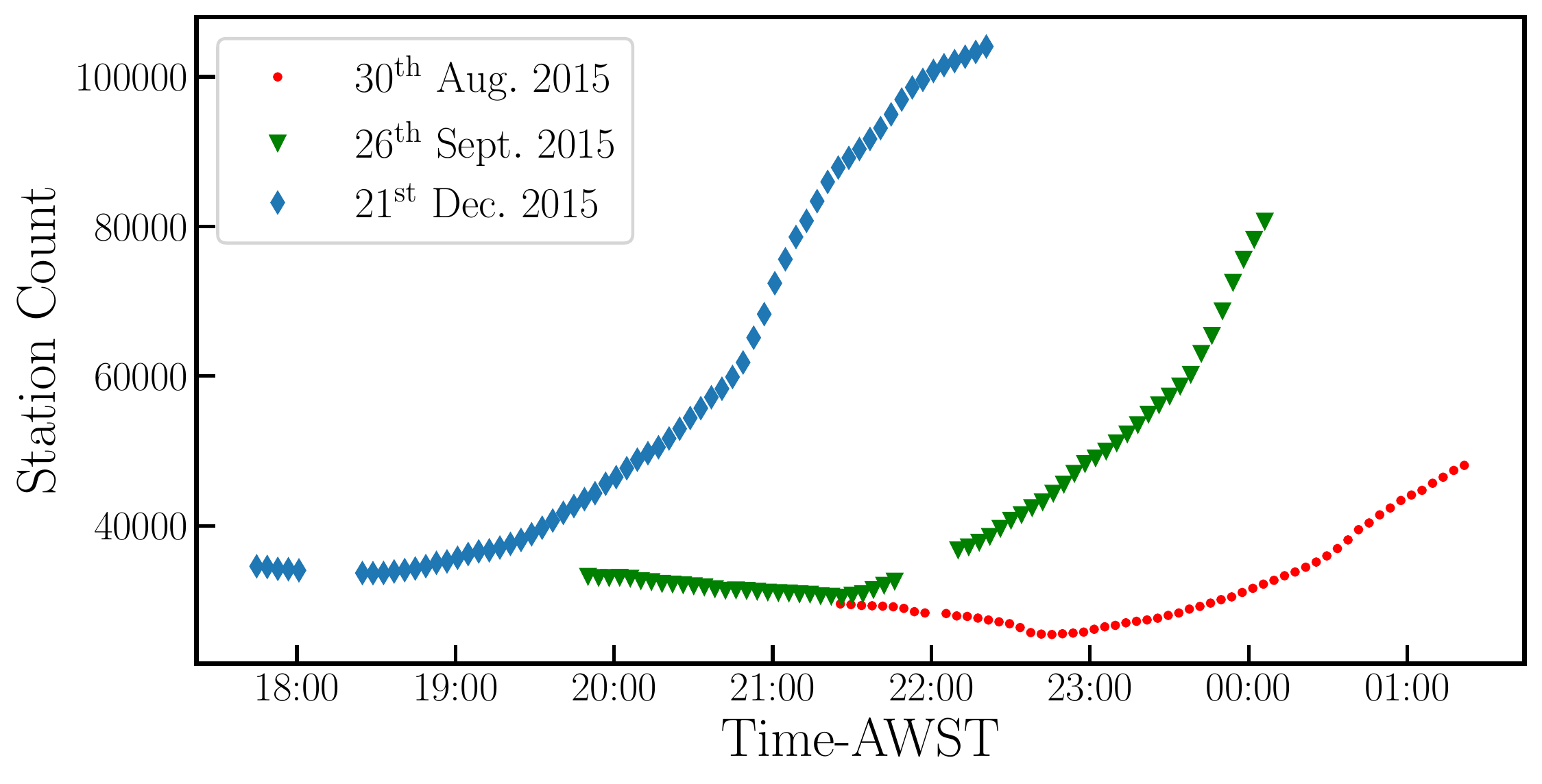}
    \caption{The variation of FM station count during the ON-Moon observations made on all three Epochs ($30^{\rm th}$ August, $26^{\rm th}$ September and $21^{\rm st}$ December 2015). It can be seen that the station counts change significantly on all three epochs during the observation. The data points are generated every $\approx 230$ seconds to match the observation time. }
    \label{fig: FM_month_sim}
\end{figure}

\section{Results} \label{section: results}

\subsection{Estimating \texorpdfstring{$S_m(\nu)$}{Lg}} \label{subsection: S_m_estimate}

In the previous steps of modelling (\S\ref{section: data_processing}), we separated out the specular component from the Moon. The removal of the remaining diffuse component $(S_{\rm diffuse})$ will provide an estimate of the flux density of the Moon $(S_m(\nu))$,
\begin{equation}
\label{eq: moon_flux}
    S_m(\nu) = S_{\rm disk}(\nu) - S_{\rm diffuse} (\nu)
\end{equation}
The reflected power of the specular and the diffuse components have a frequency dependence (see eq. 31 and 32 from \cite{Evans_1969}). Applying these equations to the radar equation (eq. 24 from \cite{Evans_1969}, or a similar equation described herein \ref{subsection: simulation_method}), one can describe a relation between specular and diffuse components as,
\begin{equation}
\label{eq: moon_diffuse}
    S_{\rm diffuse}(\nu) = R_e(\nu)  S_{\rm spec} (\nu)
\end{equation}
Due to the usefulness of the FM band in our analysis, we defined $R_e(\nu)$ for a scaling frequency $(\nu_0=100$ MHz) in the middle of the FM band, with the power-law index 0.58 arriving from the frequency dependence (see eq. 31 and 32 from \cite{Evans_1969}).
\begin{equation}\label{eq: R_e}
    R_e(\nu) = \frac{S_{\rm diffuse}(\nu_0)}{S_{\rm spec} (\nu_0)}\left(\frac{\nu}{\nu_0}\right)^{0.58}, \nu_0 = 100 \rm~ MHz
\end{equation}
Once the $S_{\rm diffuse}(\nu)$ is estimated $S_m(\nu)$ can evaluated using eq. \ref{eq: moon_flux} . Next describes the two methods we used to estimate $S_{\rm diffuse}(\nu)$ as the final Earthshine mitigation step before estimating the foreground sky temperatures.

\subsubsection{From DATA} 
\begin{figure*}[ht]
    \centering
    \includegraphics[scale=0.45]{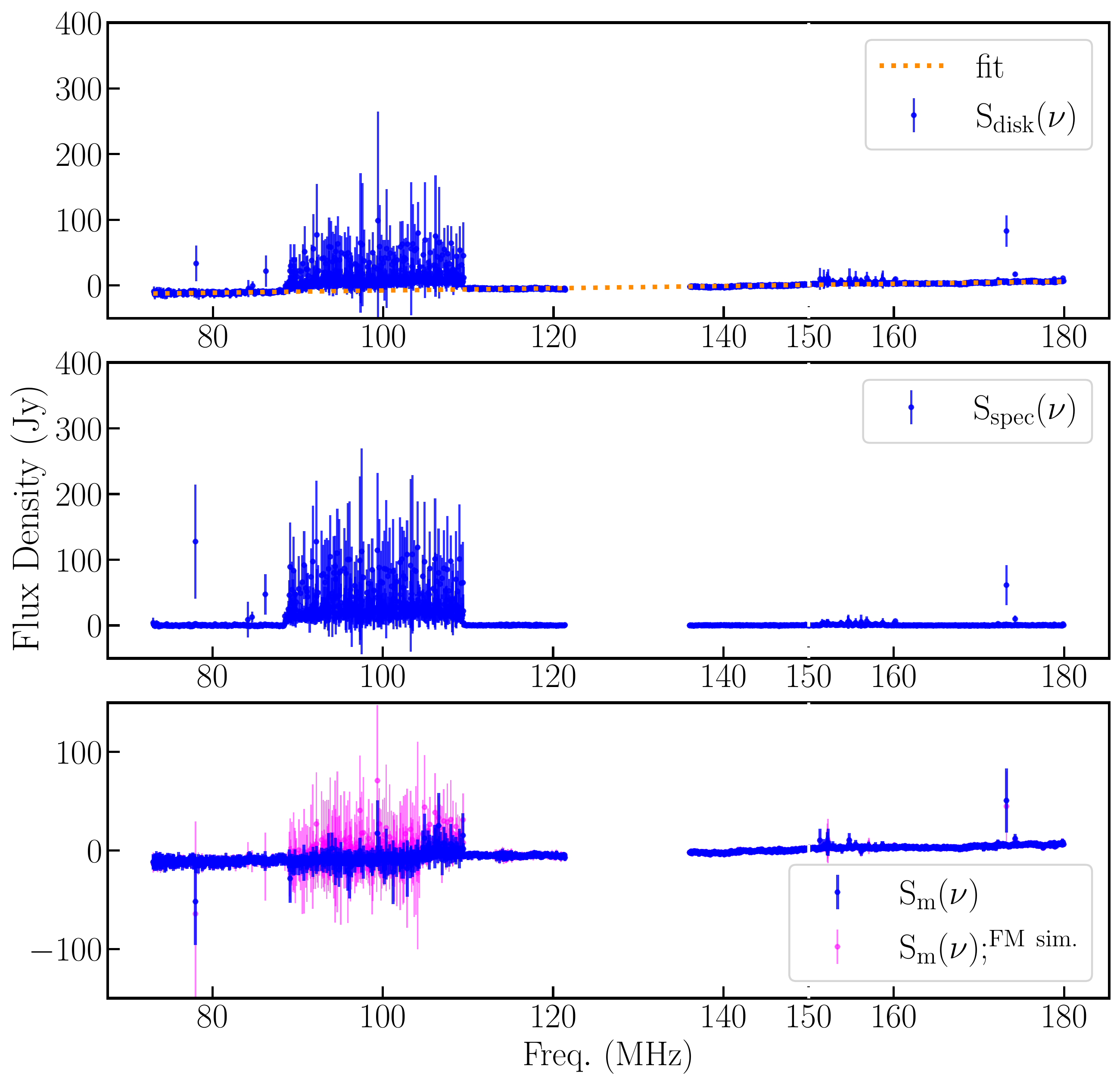}
    \caption{The observed flux density of the disk component in the top panel and the quasi-specular Earthshine component in the middle panel. The bottom panel shows the flux density of the Moon after performing the Earthshine mitigation using two methods. The \textit{blue} errorbars correspond to the Earthshine mitigation of the first kind, where we used the fitted value of $S_{\rm diffuse}(\nu)$ at $\nu_0$ to determine $S_{m}(\nu)$.  The line fitted to the disk component is shown with \textit{yellow} dashes in the top panel. The second method is shown by the \textit{pink} errorbars, where the simulated value of  $S_{\rm diffuse}(\nu)$ at $\nu_0$ is used to estimate $S_{m}(\nu)$. The data points are at 40 kHz fine channels and averaged over 12 full-band MWA observations from the first epoch. The corresponding errors are obtained using the inverse variance weighted scheme. The \textit{black} vertical dashed line is at 150 MHz, which corresponds to the frequency where the Moon first appears in emission.}
    \label{fig: Flux_density_all}
\end{figure*}

In the first method, the flux-density $S_{\rm diffuse}(\nu_0)$ is obtained by fitting a line to the $S_{\rm disk}(\nu)$ component (see fig.\ref{fig: Flux_density_all}, yellow fitted line on the top panel). The fitted value at $\nu_0$ represents $S_{m}(\nu_0)$, and rearranging eq.\ref{eq: moon_flux} can provide $S_{\rm diffuse}(\nu_0)$. Once we obtained the $R_e(\nu)$, we plugged it into eq.\ref{eq: moon_diffuse} and determined  $S_{\rm diffuse}(\nu)$ for the entire frequency band, and finally we remove the Earthshine component $S_{\rm diffuse}(\nu)$ from the  $S_{\rm disk}(\nu)$ and estimated $S_m(\nu)$ using eq. \ref{eq: moon_flux}.

\subsubsection{From Simulation}\label{subsection: simulation_result}

In the Earthshine simulations, we first calculated the number of FM stations from where the Moon was above the horizon. This was used to estimate the total RFI reflection from the disk of the Moon. It can be seen (see fig. \ref{fig: flux_vs_st_count}) that the reflected flux density received at the MWA increases with increasing station count. For an entire ON-Moon observation epoch, we simulated the Earthshine for every ON-Moon observation (see the variation of FM station counts as it changes over the entire observation duration of $\approx 3$ hours in fig. \ref{fig: FM_month_sim}). However, in our analysis, we used the simulated Earthshine only at those timestamps when our ON-Moon observations were in the FM band (i.e. when the 30.72 MHz contiguous band included the FM frequencies).  
The simulated Earthshine flux density can be used to replace the flux density from the data in the FM band (here, we refer to $S_{\rm diffuse}$ between $88-110$ MHz as the data which were evaluated using eq. \ref{eq: moon_diffuse}).

\begin{figure}[ht]
    \centering
    \includegraphics[scale=0.50]{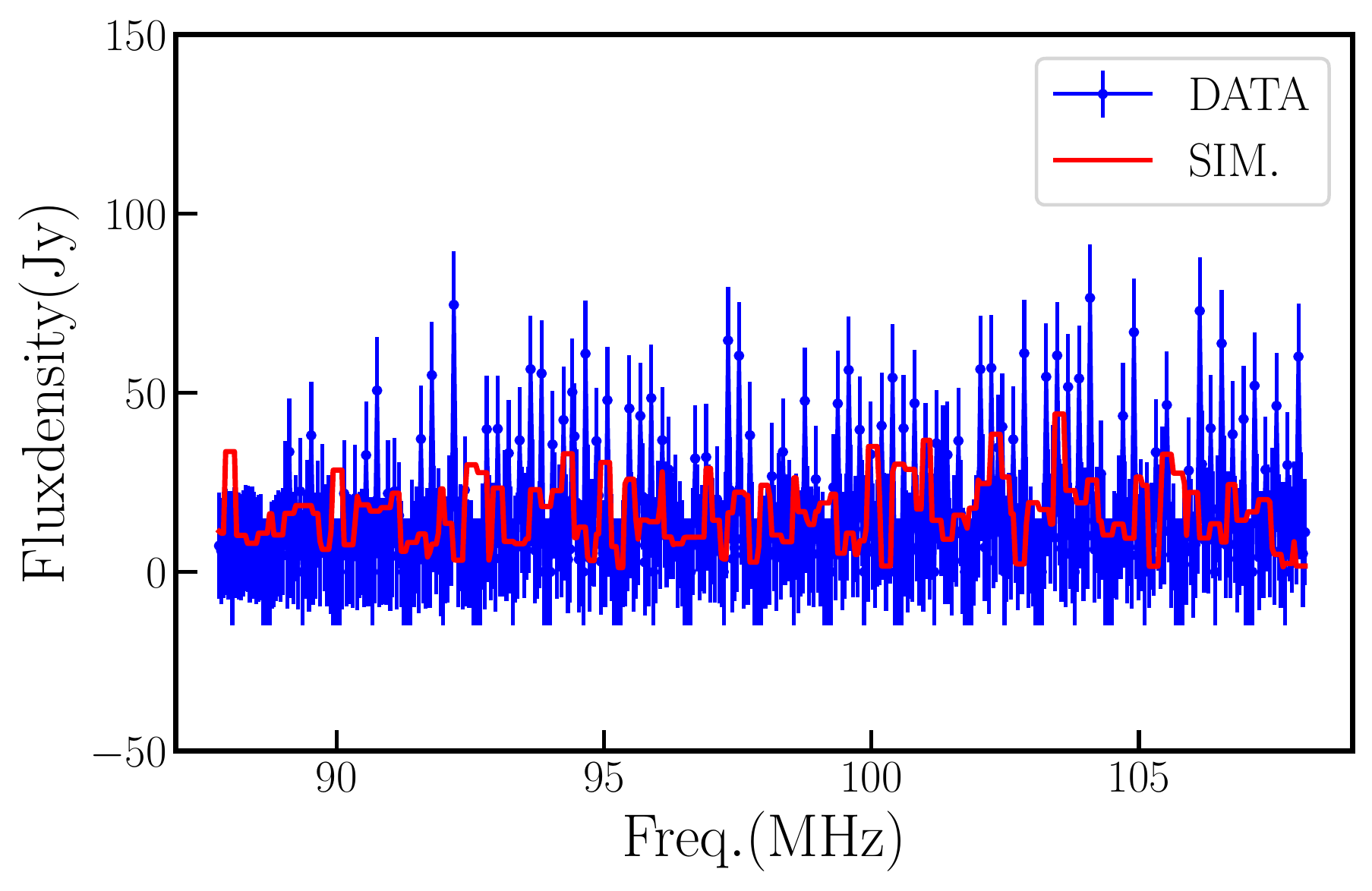}
    \caption{The mean flux density of the data and simulation in the FM band at the first observing epoch.}
    \label{fig: data_vs_sim_flux}
\end{figure}

To statistically quantify the simulations and the data we performed a simple T-test on the data and the simulation. Our null hypothesis was based on the argument that the two discrete samples are drawn from the same distribution. We used the T-test with a confidence of $95\%$, which means we rejected the null hypothesis if the $p$-values were less than 0.05 and accepted otherwise. Our sample set comprised the LST variation (across a given observing epoch) of the flux density of the data and simulation with the frequency resolution of 40 kHz and matching LST cadence to the observing epochs (note that while estimating the simulated flux density we did not account for the integration time but rather estimated the instantaneous flux density at the middle of each observation). The $S_{\rm diffuse}$ from the data is obtained by performing a similar fitting (mentioned as the first approach) to $S_{\rm disk}$ for all 34 full-band individually which also provides the LST variation of $S_{\rm diffuse}$.
Once the flux density $S_{\rm diffuse}, S_{\rm FM}$ for every individual epoch is evaluated we performed the T-test between them at every $40$ kHz fine channel independently. The estimated $p$-values on the dataset from the first epoch are shown in fig. \ref{fig: p_value_plot}. As we are limited by our FM catalogue and simulations, our null hypothesis was rejected in nearly half of the fine frequency channels. We note that the simulations were generated at the frequency resolution of the data (i.e. 40 kHz) with the assumption that the FM stations transmit constant power throughout the 180 kHz bandwidth, so a constant flux density was estimated by the simulations, which is not true in case of the data (see fig. \ref{fig: data_vs_sim_flux} showing the flux-density of the data $(S_{\rm diffuse})$ and simulation $(S_{\rm FM})$). 
Therefore, instead of directly subtracting the reflected diffuse FM RFI component from the data, we restricted our analysis to the single frequency channel, in particular, the central FM band $(\approx \nu_0)$ and evaluated the $R_e(\nu)$ (from eq.\ref{eq: R_e}) to perform the Earthshine mitigation. The estimated $p$-values at the middle of the FM band $(\approx \nu_0)$ are presented in Table \ref{tab: T-test}.
\begin{table}[ht]
    \centering
    
    \begin{tabular}{c|c}
    ${\rm Epochs}$ & $p$-value at $(\approx \nu_0)$ \\
    \hline
    ${\rm Aug.}$ &  0.75 \\
    ${\rm Sept.}$ &  0.69 \\
    ${\rm Nov.}$ &  0.78  

    \caption{$p$-values from the T-test at the middle of the FM band $\approx \nu_0$.}
 
    \label{tab: T-test}
    \end{tabular}
\end{table}

Since the $p$-values at $\nu_0$ satisfy our null hypothesis, in the second method of Earthshine mitigation, we used the simulated value of $S_{\rm FM}(\nu_0)$ as the $S_{\rm diffuse}(\nu_0)$ and estimated $R_e(\nu)$ and used it in eq.\ref{eq: moon_diffuse}  and eq.\ref{eq: moon_flux} to determine $S_m(\nu)$. The flux density of the Moon obtained via both methods is shown in fig.\ref{fig: Flux_density_all}.  

In both methods of Earthshine mitigation, we obtained the values of $S_m(\nu)$ for every 34 full-band observations separately, and to obtain the uncertainties in $\Delta S_m(\nu)$. We used the RMS noise of the disk and quasi-specular components, the analysis of which is presented in the appendix. Finally, we used the inverse-variance weights to obtain the mean and variance in $S_m(\nu)$. We averaged $S_m(\nu)$ according to their corresponding epochs. As a result, we had 3 realisations of $S_m(\nu)$ for the corresponding 3 ON-Moon observing nights which were used in eq.\ref{eq: delta_t} to find the temperature difference $\Delta T(\nu)$.

\subsection{Estimating \texorpdfstring{$T_{\rm refl-Gal}(\nu)$}{Lg}}

\begin{figure*}[ht]
    \includegraphics[scale=0.56]{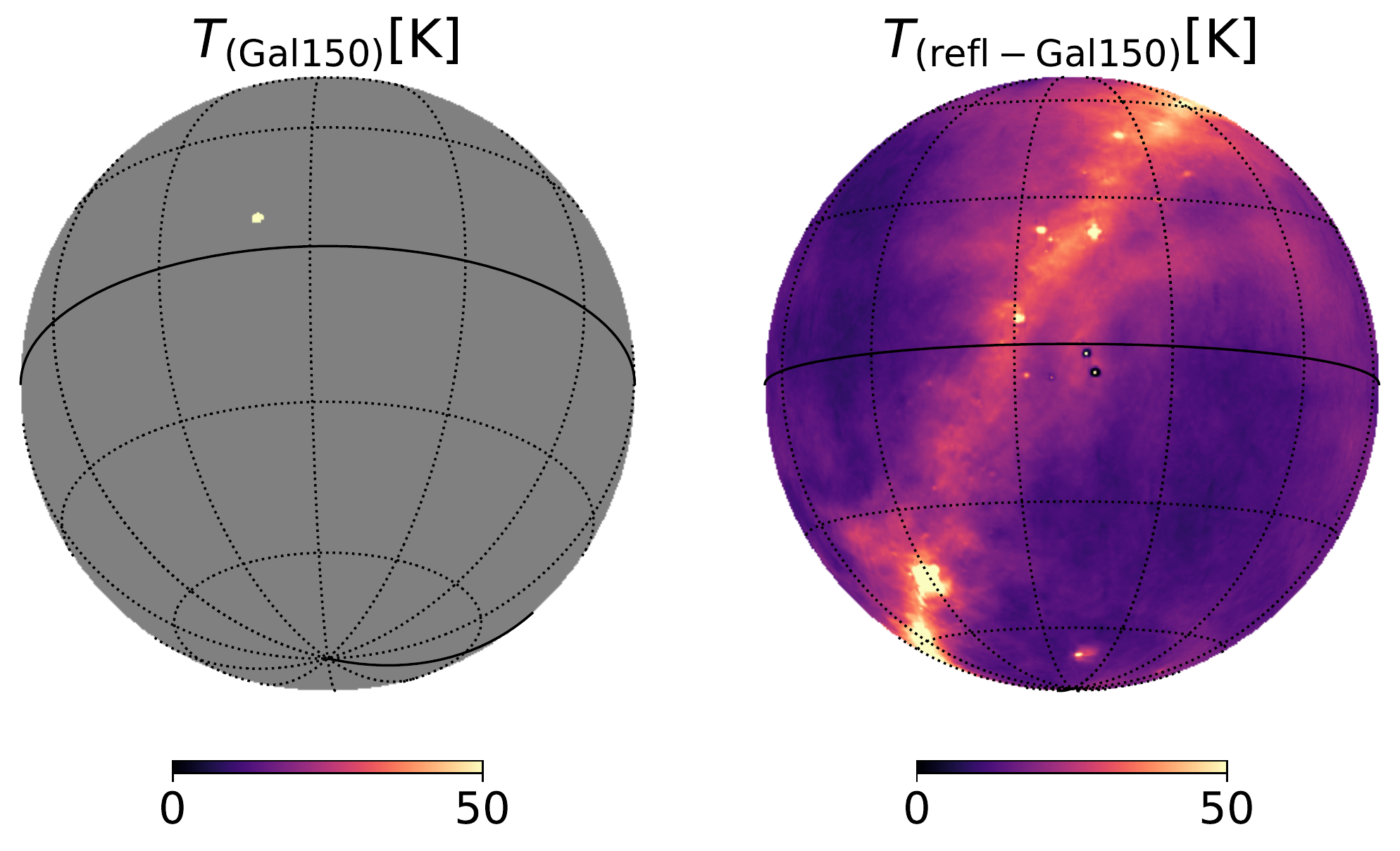}
    \caption{Galactic temperature from the GSM2016 model at the middle of the second epoch. The left panel shows the occulted sky temperature $T_{\rm Gal150}$ as observed from the MWA. The single dot represents the location of the Moon during the ON-Moon observation. The right panel shows the sky temperature reflected by the Moon $T_{\rm refl-Gal}$. To make the pixels visible on the left figure, we doubled the pixel counts of the Moon and saturated the colourbar by using the same colour scale as the right figure.}
    \label{fig: ref_gal_tmp_map}
\end{figure*}

We used a similar vector ray-tracing algorithm to \citet{Ben_2018} to estimate the reflected Galactic emission from the Moon. The algorithm assumes that the observatory is located on the Moon and generates the sky map as observed from the Moon. The part of the sky map that reflects back to MWA's location is determined by estimating the angle of incidence and reflection of the sky map at the Moon's surface based on the radar cross-section criteria. We used Python-based \texttt{PYGDSM} to produce the Low-Frequency Survey Model (LFSM) \citep{LWA1_LFSM_Dowell}, Haslam \citep{Haslam_Remazeilles}, Global Sky Model 2008 \citep{GSM2008_de_Costa}, Global Sky Model 2016 \citep{Zheng_2016} models for all three ON-Moon observation epochs. The sky models were then injected into our algorithm to produce the reflected sky map from the Moon. The reflected sky does not change significantly during our observational epochs (spanning approx $2-3~$hours), as the Moon sweeps only about $(\approx 1^\circ-1.5^\circ )$ on the sky during each epoch. Therefore, we used the middle point of time at each MWA's full band observations to estimate the reflected Galactic temperature. We produced the sky maps at every $5$ MHz frequency channel from $70-180$ MHz. Fig. \ref{fig: ref_gal_tmp_map} shows the reflected Galactic temperature $T_{\rm refl-Gal}$ at $150$ MHz in the middle of the first ON-Moon observation epoch. 
$T_{\rm refl-Gal}(\nu)$ follows a power law, and it can be shown that the reflected Galactic temperature fits well with the power-law equation:
\begin{equation}\label{eq: refl_temp}
    T_{\rm refl-Gal}(\nu) = T_{\rm refl-Gal150} \left(\frac{\nu}{150 \rm MHz}\right)^\beta
\end{equation}
\begin{figure}[ht]
    \centering
    \includegraphics[scale=0.36]{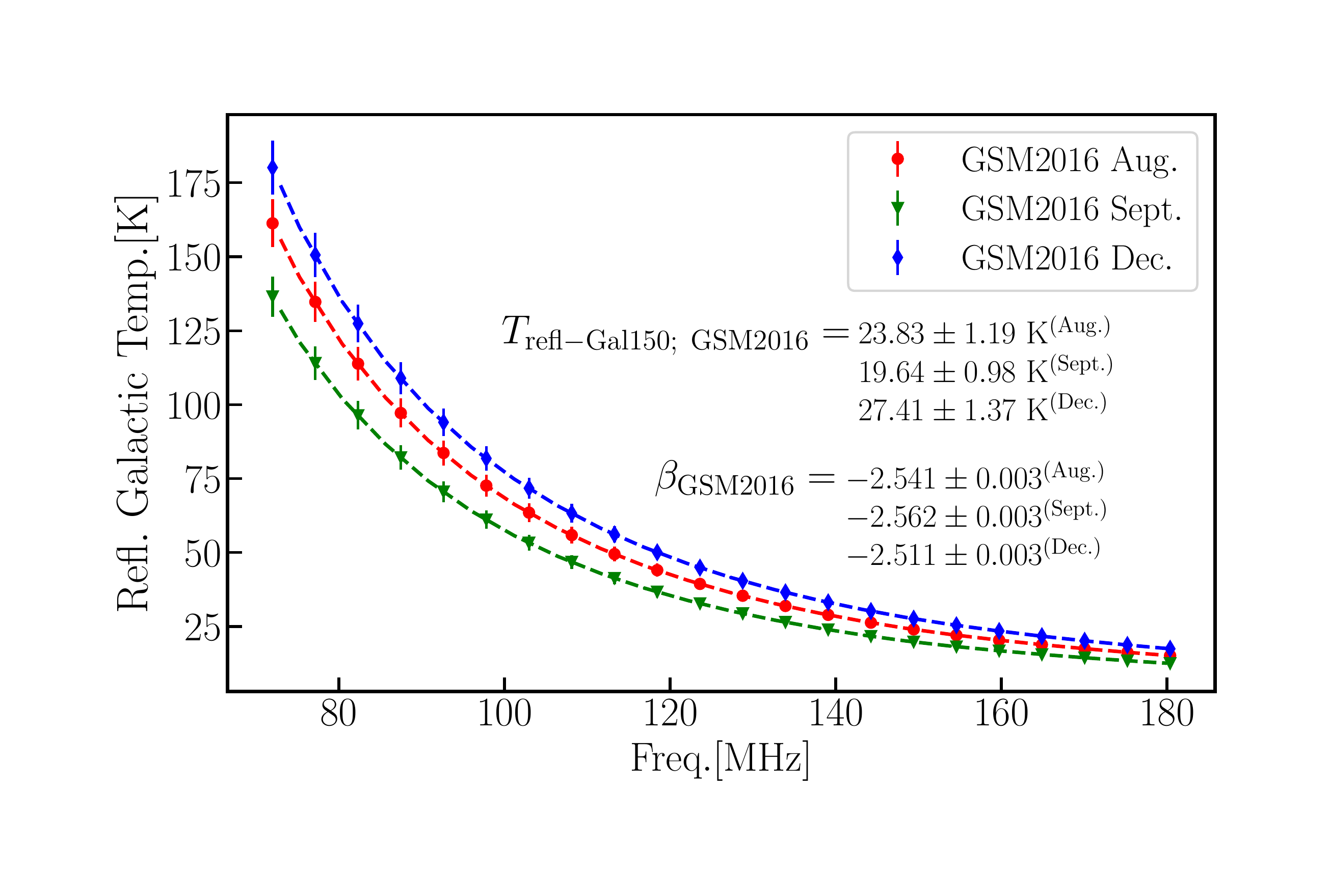}
    \caption{Mean reflected Galactic temperature estimated using GSM2016 \citep{Zheng_2016} at all three epochs. The reflected Galactic temperature (shown by point, triangle, and diamond markers) is fitted with the power law eq.\ref{eq: refl_temp}. It can be seen that the reflected Galactic temperature at $150$ MHz $(T_{\rm refl-Gal150})$ does not change significantly between the epochs. The error bars corresponding to $T_{\rm refl-Gal150}$ represent the $5\%$ model estimation error of GSM2016\citep{Zheng_2016}, and the uncertainty in the $\beta$ represents the fitting error.}
\label{figure: ref_gal_tmp}
\end{figure}

The spectral index $\beta$ is obtained by fitting the mean and $5\%$ model error \citep{Zheng_2016} of the reflected Galactic temperature to the power-law equation. Fig.\ref{figure: ref_gal_tmp} shows the reflected Galactic temperature for all of the three epochs.  The fitted values of  $T_{\rm refl-Gal150}$ and $\beta$, are provided in Table \ref{tab: Refl_Gal_Temp_table}.

\begin{table}[ht]
    \centering
    \begin{tabular}{c|c|c}
    ${\rm Epochs}$ & $T_{\{\rm refl-Gal150;~ GSM2016\}}$ & $\beta_{\rm GSM2016}$\\
    \hline
    ${\rm Aug.}$ & $23.83\pm{1.19}\rm K$ & $-2.541\pm{0.003}$  \\
    ${\rm Sept.}$ &  $19.64\pm{0.98}\rm K$ & $-2.562\pm{0.003}$  \\
    ${\rm Nov.}$ & $27.41\pm{1.37}\rm K$ & $-2.511\pm{0.003}$
    \caption{Showing the fitted values of the $T_{\rm refl-Gal150}$ and reflected Galactic spectral index $\beta$.}
    \label{tab: Refl_Gal_Temp_table}
    \end{tabular}
\end{table}
   
\subsection{Estimating \texorpdfstring{$T_{\rm Moon}$}{Lg} and \texorpdfstring{$T_{\rm Gal}(\nu)$}{Lg}}

So far, we have evaluated the quantities $\Delta T$, $T_{\rm refl-Earth}$ and $T_{\rm refl-Gal}$ of eq.\ref{eq: temp_all}. Note that the $T_{\rm refl-Earth}$ has already been removed from the data during the Earthshine mitigation process (described in the previous section \S\ref{subsection: S_m_estimate}). Therefore, we are left with the $T_{\rm Gal}(\nu)$, $T_{\rm Moon}(\nu)$, $T_{\rm EoR}(\nu)$ and $T_{\rm CMB}$ variables. Our present analysis focuses on combining the observations from different epochs and checking whether we can produce better constraints on the Moon's intrinsic temperature from our previous work. Detecting the EoR would require proper foreground modelling, more observations to increase the SNR of the occulted sky patch and improved Earthshine models, which we aim to address in future works. Therefore, we ignore the contribution of EoR in the sky temperature. 
The remaining variables $T_{\rm Gal}(\nu)$ and $T_{\rm Moon}$, are the temperature of the occulted patch of sky and the intrinsic temperature of the Moon, respectively. The variables in eq.\ref{eq: delta_t} can be rearranged to:
\begin{equation} \label{eq: T_gal-moon_data}
     T_{\rm Gal}(\nu) - T_{\rm Moon} =
    T_{\rm refl-Gal}(\nu) - \Delta T (\nu) - T_{\rm CMB}
\end{equation}
Here we have information on the RHS variables of the equation, and as we considered the Moon to be at a constant temperature, it will act as a temperature offset to the LHS. We fit the temperature difference between the Galactic emission and Moon with a similar power law from eq. \ref{eq: refl_temp}, but with a constant temperature offset $(T_{\rm offset})$.
\begin{equation}\label{eq: T_gal_moon_fit}
   T_{\rm Gal}(\nu) - T_{\rm Moon} =
     T_{\rm Gal150} \left(\frac{\nu}{150 \rm MHz}\right)^\alpha - T_{\rm offset}
\end{equation}
\begin{figure*}[ht]
    \includegraphics[scale=0.5]{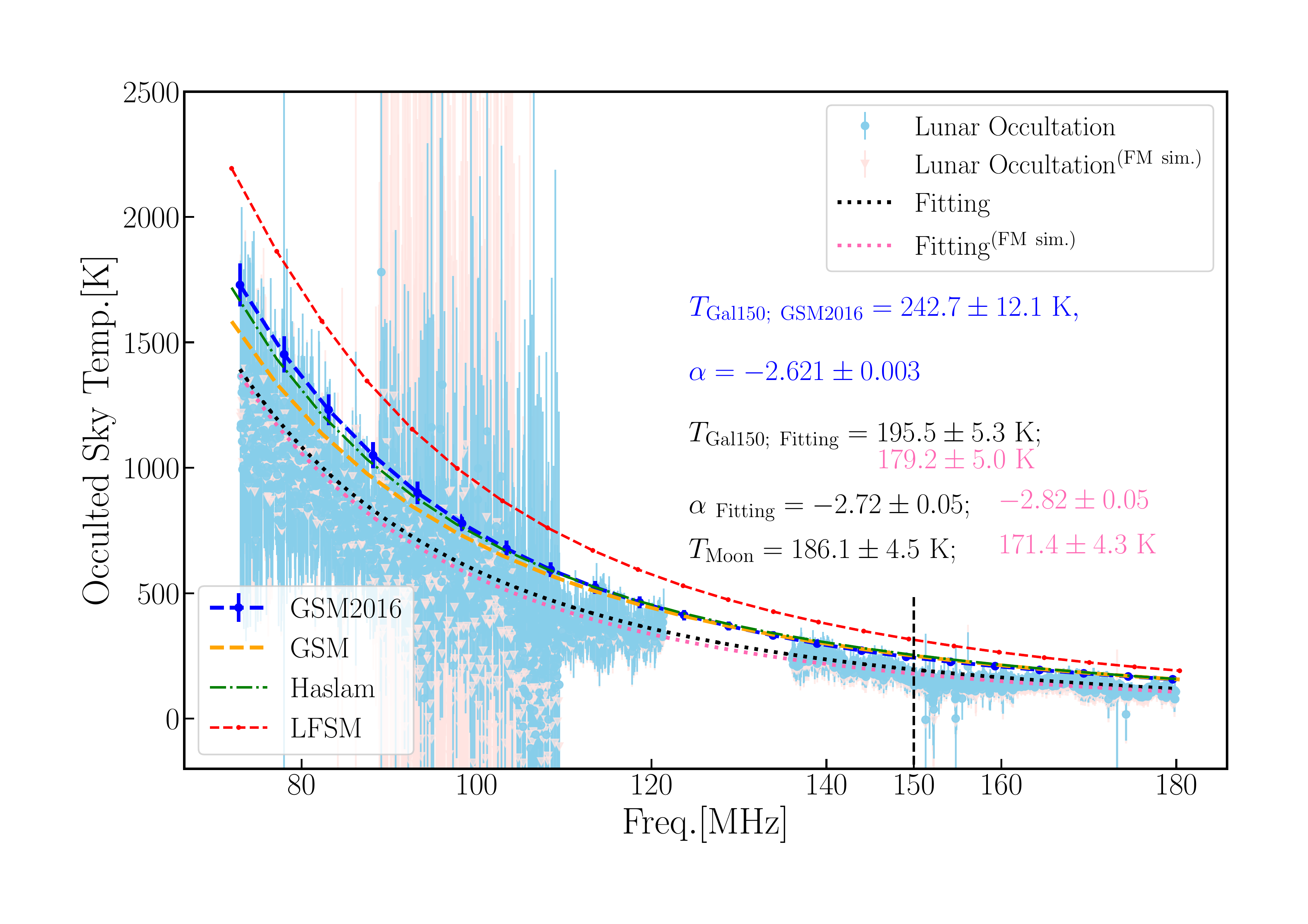}
    \caption{Occulted sky temperature from the dataset from the first epoch. The data points shown by \textit{light blue} and \textit{light pink} represent the occulted sky temperature obtained using the two Earthshine mitigation methods. The data points fitted with $T_{\rm Gal}(\nu)$, are shown by the dashed lines (\textit{black} and \textit{pink}) for the respective methods.  The quoted values of the $T_{\rm Gal150;~ Fitting}$, $\alpha_{\rm Fitting}$ and $T_{\rm Moon}$ are obtained by taking the inverse-variance weighted mean and variance. For comparison, different GSM sky models are plotted over the data points. The model uncertainty in the Galactic temperature and spectral index of GSM2016 are shown in the \textit{blue} text in the figure.}
    \label{fig: gal_temp_with_FM_rem_epoch1}
\end{figure*}

\subsubsection{Fitting for \texorpdfstring{$T_{\rm Moon}$ (Individual Epochs)}{Lg}}

    In order to obtain $T_{\rm Moon}$, first, we estimated the model values of the $T_{\rm Gal}(\nu)$, and we followed the same procedure as for $T_{\rm ref-Gal}(\nu)$. We generated the GSM, GSM2016, LFSM and Haslam maps at the frequency resolution of 5 MHz for all ON-Moon observations and estimated the $T_{\rm Gal}(\nu)$ of the patch of sky occulted by the Moon. \\We fit the model Galactic temperature with a similar power-law equation to $T_{\rm refl-Gal}$  (see \S\ref{app: Tgal_Tref_gal_err} and eq.\ref{eq: T_gal}, and obtained the values of $T_{\rm Gal150;~sky-model}$, and $\alpha_{\rm Gal150;~sky-model}$. 
    In fig.\ref{fig: gal_temp_with_FM_rem_epoch1}, we show the data from the first epoch. The $T_{\rm Gal}(\nu)$ is estimated using both methods of Earthshine mitigation, and the fitted values of $T_{\rm Gal}(\nu)$ at 150 MHz and the Galactic spectral index $(\alpha)$ are presented for the data and GSM2016 model. A comprehensive table showing the fitted values of $T_{\rm Gal}(\nu=150 \rm ~MHz)$ and $T_{\rm Moon}$ estimates from all sky models is given in the appendix (see \ref{tab: table_individual_all_models}). Please note that, in order to get the fitting for $T_{\rm Moon}$ and $T_{\rm Gal150}$ for a given sky model, we first estimated the reflected sky temperature from the same sky model and repropagated it through eq. \ref{eq: T_gal-moon_data} and \ref{eq: T_gal_moon_fit}.
    
\subsubsection{Joint Fitting for \texorpdfstring{$T_{\rm Moon}$}{Lg}}
 
\begin{figure*}[ht]
    \centering
    \includegraphics[scale=0.45]{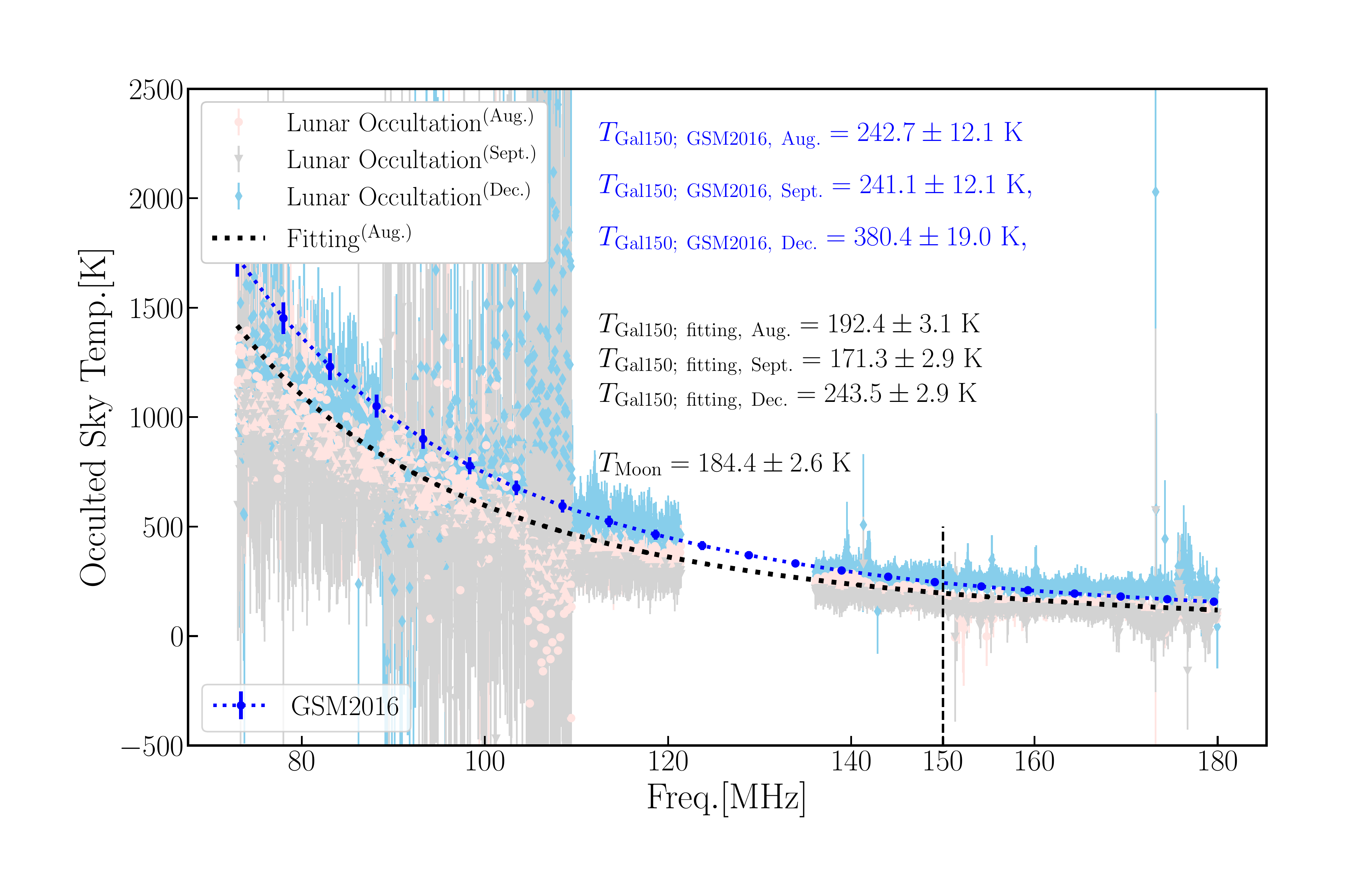}
    
    \caption{The data points from the three observational epochs are shown with different colour schemes (\textit{light blue, light grey, light pink}). The \textit{black} dashed line corresponds to the best fit to the occulted sky temperature from the first epoch (August 2015) and is plotted along with the \textit{blue} errorbars of the GSM2016. The data points and fitted values are the estimates obtained using the first method of Earthshine mitigation. }
    \label{fig: T_gal_all_method1}
\end{figure*}

\begin{figure*}[ht]
    \centering
    \includegraphics[scale=0.45]{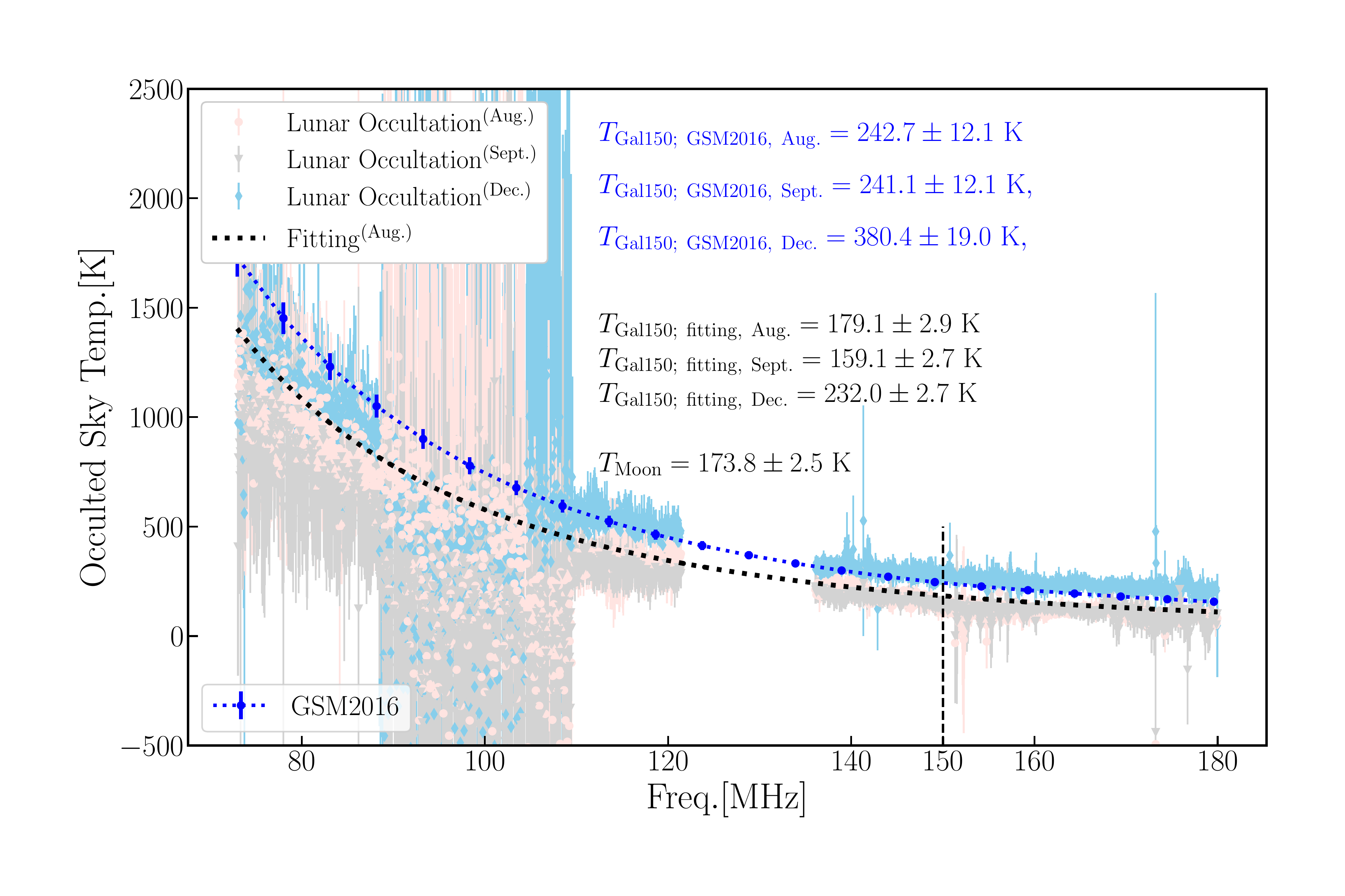}
    
    \caption{Occulted sky temperature measured using the second method of Earthshine mitigation (FM simulation). Data points from all three epochs are presented with coloured errorbars. The best fit to the first epoch (August 2015) dataset is plotted with the black dashed line, and the corresponding values of the GSM2016 model are plotted with the blue errorbars.}
    \label{fig: T_gal_all_method2}
\end{figure*}

We combined the measured values of $\Delta T (\nu)$ from all three epochs and performed a joint fit to eq.\ref{eq: T_gal_moon_fit}. We placed restrictions on the $T_{\rm Moon}$ and considered the Moon to have a constant temperature. In the joint fit of the data to eq.\ref{eq: T_gal_moon_fit}, there is a total of six independent and one dependent parameter. The best-fit values of $T_{\rm Gal150}$, spectral index $\alpha$ and  $T_{\rm Moon}$, along with the estimations of the sky models, are presented in Table \ref{tab: table_joint_results}. The results presented in Table \ref{tab: table_joint_results} are for both the Earthshine mitigation approaches. The table also shows the expected Galactic foreground temperature and the spectral index (obtained from the GSM2016 model \citep{Zheng_2016}) of the patch of sky occulted by the Moon during these observations.  Please note that the larger error bars in fig.(\ref{fig: gal_temp_with_FM_rem_epoch1}, \ref{fig: T_gal_all_method1}, \ref{fig: T_gal_all_method2}) at the lower frequencies could arise due to several factors, the first being the observations itself. Since the Moon moves  $\approx 6.5^\circ$ with respect to the background sky during a single night, the mean occulted galactic temperature obtained over multiple individual spectra shows a larger deviation. Second, the thermal noise and the noise from the sidelobe confusion are more at the lower frequencies. In general, the sky noise is dominated by the lower frequencies.
\begin{table*}
    \centering
    \begin{tabular}{c|c|c|c|c|c}
    \toprule
    \multicolumn{6}{c}{$\rm Method~ 1$}\\
    \hline
     $\rm Epochs$  & $T_{\{\rm Gal150;~ GSM2016\}}\rm ~(K)$ & $\alpha_{\rm GSM2016}$ & $T_{\{\rm Gal150;~fitting\}}\rm(K)$ &  $\alpha_{\rm fitting}$  &  $T_{\{\rm Moon\}}\rm(K)$  \\
        \midrule
        $\rm Aug.$ &  $242.7\pm{12.1}$ &  $-2.621 \pm{0.003}$ & $192.4\pm{3.1}$ & $-2.745\pm{0.031}$ & \\
        $\rm Sept.$  & $241.0\pm{12.0}$& $-2.585\pm{0.002}$ & $171.3\pm{2.8}$& $-2.598\pm{0.033}$ & $184.4\pm{2.6}$ \\
$\rm Dec.$ & $380.4\pm{19.0}$& $-2.497\pm{0.002}$ & $243.5\pm{2.9}$  & $-2.612\pm{0.022}$     & \\
        \hline
        \multicolumn{6}{c}{$\rm Method~ 2^{\rm ~(FM~sim.)}$}\\
        \hline
        $\rm Aug.$ &  $242.7\pm{12.1}$ &  $-2.621 \pm{0.003}$ & 
 $179.1\pm{2.9}$ & $-2.798\pm{0.033}$ & \\
        $\rm Sept.$ &  $241.0\pm{12.0}$& $-2.585\pm{0.002}$ & $159.1\pm{2.7}$  &  $-2.640\pm{0.034}$ & $173.8\pm{2.5}$\\
        $\rm Dec.$ &  $380.4\pm{19.0}$ & $-2.497\pm{0.002}$ & $232.0\pm{2.7}$ & $-2.661\pm{0.021}$ &\\
    \bottomrule
    \end{tabular}
    \caption{Table showing the best-fit parameters from the joint-fitting of the combined epochs.}
    \label{tab: table_joint_results}
\end{table*}

\section{Discussion}  \label{section: discussion}

As the $T_{\rm Gal}(\nu)$, $T_{\rm Moon}$ and the spectral index $\alpha$ are deduced from the same fitting function, these parameters are highly correlated with each other. The correlation amongst the parameters from the joint analysis for the first method of  Earthshine mitigation is shown in Table \ref{tab: joint_corr_table}.
\begin{figure}[ht]
    \centering
    \caption{Measurement of the $T_{\rm Moon}$ from our analysis along with the \citet{Ben_2018} result and Table 2 from \citet{Krotikov_1964}. The values of $T_{\rm Moon}$ from this work are presented as points at $150$ MHz.}
    \includegraphics[scale=0.40]{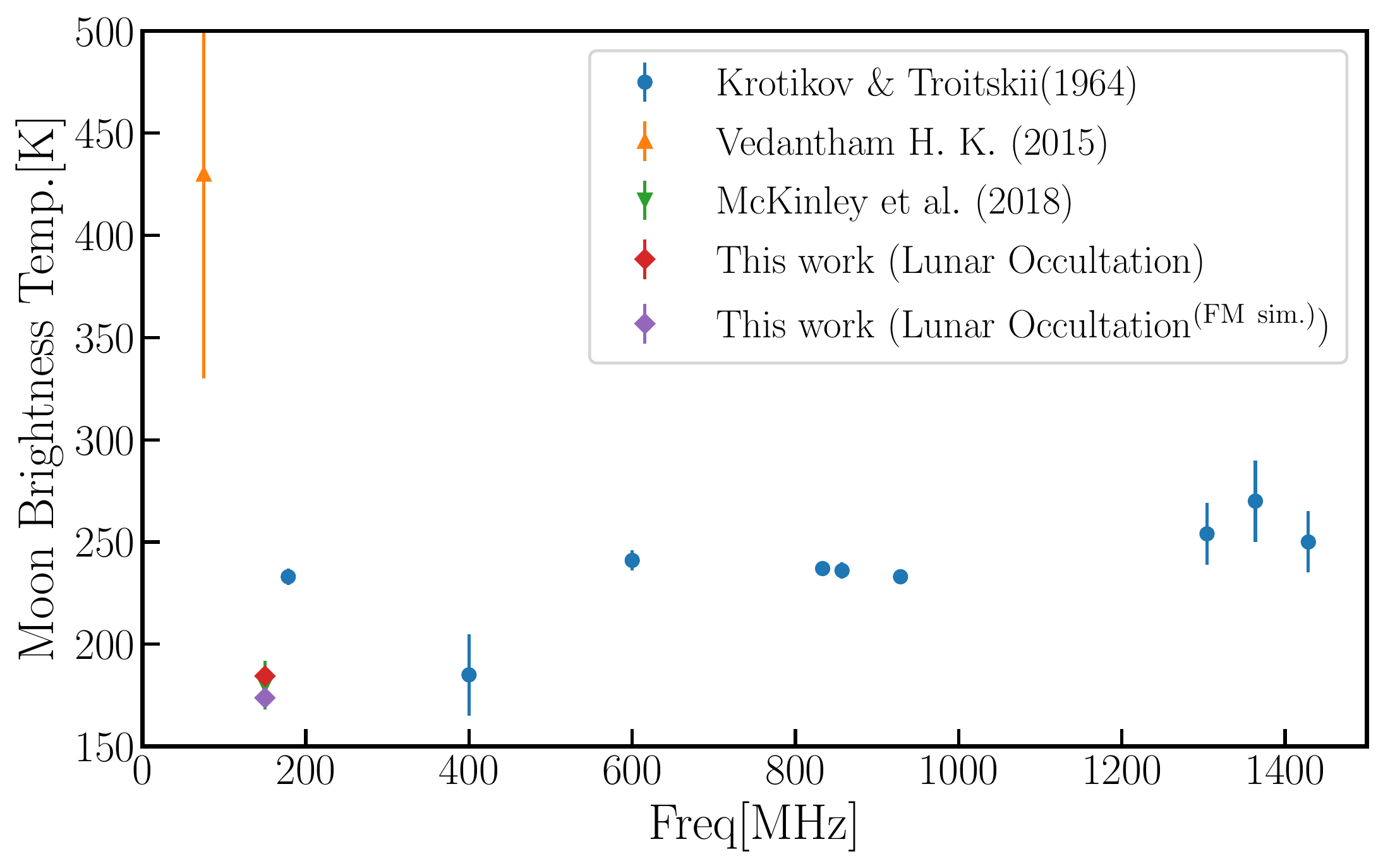}
    \label{fig: T_moon_his}
\end{figure}
We aim to deal with these degeneracies in future work when we include the observational datasets from the Engineering Development Array (EDA) \citep{wayth_2017} and measure the Moon temperature independently, which can provide better constraints on the Galactic foregrounds. Our finding of the Moon's temperature, along with \citet{Ben_2018} and results from Table $2$ of \citet{Krotikov_1964}, are shown in fig.\ref{fig: T_moon_his}. 
In our joint analysis, we obtained the Moon temperature from the first method to be $184.4\pm{2.6} \rm K$ and $173.8\pm{2.5} \rm K$ from the second method, respectively.  These estimates are inconsistent with each other. The deviation in $T_{\rm moon}$ between the two methods is $\approx 10\rm K$, which could be due to imperfect simulations; however, these results put tighter constraints on the Moon temperature and agree with the results of \citet{Ben_2018}, which predicted $T_{\rm moon}= 180\pm{12}\rm K$. 
 On the other hand, similar to \citet{Ben_2018},  our estimates of the occulted sky temperature at 150 MHz ($T_{\rm Gal150}$) underestimated the model predictions. Our results agree within $2\sigma$ uncertainty levels when compared with the GSM2016 sky model \citep{Zheng_2016} (note that here we used the $\sigma$ levels by considering the uncertainties as Gaussian around the mean predicted values of $T_{\rm Gal150}$). 
 
 When we compared our best-case scenario for a single epoch with the GSM2016 model, our estimates of  $T_{\rm Gal150}$ had $\approx 19\%$ error with the model predictions. In comparison, the same was $\approx 12\%$ in \citet{Ben_2018} results. It can be argued that our frequency resolution of 40 kHz, which is higher compared to  \citet{Ben_2018} (1.28 MHz), could have provided excess noise in the flux density estimation. 
The Galactic spectral index $(\alpha)$, when compared to the corresponding GSM2016 predicted values, agrees within $5\%$ uncertainty levels in all three observation scenarios in the first method and $7\%$ uncertainty in the second method. The spectral index measured at all three epochs is consistent with the findings of \citet{Ben_2018}. When comparing the spectral index from other global sky models (see. fig.\ref{fig: gal_temp_with_FM_rem_epoch1}),  our results agreed, except for the LFSM, which predicted a higher steepness. We note that the spectral index changes from one part of the sky to another. 

We compared our joint-fit results with the individual fits to the epochs. Since the three epochs are at different LSTs, one can argue how the joint fit will affect the  $T_{\rm Gal150}$ estimates of individual epochs. We see that the deviation in the mean temperature estimates from the individual epoch $T_{\rm Gal150}$ to the jointly fitted values is only about $2-3$K (about 1-2 per cent) and agrees with the estimated uncertainties. A comprehensive table  \ref{tab: table_individual_all_models} in the appendix section shows the individual fitting cases to all models.
We also compared the joint $T_{\rm Moon}$ estimates from all the sky models. The variation in the joint $T_{\rm Moon}$ estimates is about $1-2$ K between different models. Although all cases are consistent with each other, the GSM2016 model provides the best estimates of the uncertainties. Table \ref{tab: table_joint_all_models} (Left) shows the $T_{\rm Moon}$ estimates from all sky models.
Since the Earthshine significantly contaminates the FM band, a rectification of the data, discarding the FM and other RFI-affected bands, can be made to check whether it improves our estimates or not. Although the FM frequencies are expected to cover a crucial portion of the CD-EoR phase transition, we compared the estimates of full-band with the FM-removed cases as a test case scenario for the GSM2016 sky model. We found that for the individual epochs, the fitting uncertainties on the Galactic Temperature and Moon Temperature have variations of only about $1-2$ K, while the mean temperature varies from $1-10$ K. Also, the joint estimate of the same yields similar levels of uncertainty. The constraints from full-band and test-case can be considered up to a similar level since both are consistent within the estimated uncertainty. However, it can be seen that discarding the FM band neither improves the estimates and nor significantly underestimates the occulted sky temperature. Table \ref{tab: table_joint_all_models} (Right) shows the fitted values of $T_{\rm Moon}$ and $T_{Gal150}$ for the FM-removed scenario.

At present, we used three different nights of MWA-phase I data. However, it would be valuable the see whether the same method can be used to assimilate multiple nights and perform foreground mitigations. We leave this to future works, where we aim to include more observations from different observing nights and with varying Moon elevations with respect to the Galactic plane.

\subsection{Limitations} \label{subsection: limitations}

In this part, we address various limitations of our present work, which can be dealt with in future works. Our present work includes MWA-phase I observations from multiple nights. However, in the future, the first requirement would be to check the nature of the observations before incorporating multiple datasets into the data processing. The location of the Moon in the sky during the observation can significantly alter the quality and usefulness of the data. It would be preferred to have the Moon situated near the zenith during the ON-Moon observation (if the analysis is done based on the image plane). One limitation that we faced during our data processing was the drift and shift method of the MWA observation. As the Moon was not actively tracked by MWA, it resulted in the beam-former settings creating the beam slightly away from the location of the Moon. As a result, we had to correct the beam response in the images and perform an additional rectification on the data. We chose only datasets where the beam response was greater than that at FWHM, which limited the total integration time. The EoR signal is significantly weak compared to the foregrounds; therefore, one must deal with precise calibration of the foreground, which can create huge issues if done incorrectly. The presence of bright foreground or some missing sources in the main or side-lobes can completely contaminate the images. Therefore, better and more accurate sky models are required to get good calibration results. Also, there is a possibility that the Moon can obscure a bright foreground source during some ON-Moon observations, which can result in calibration anomalies; therefore, a careful selection of the observation is required prior to data processing. We see that the angular size of the specular RFI (number of pixels acquired by the specular reflection) at a given frequency does vary with LST. Since we used the same specular RFI mask in the modelling process, it underestimated the quasi-specular component. LSTs having a wider angular size of the specular reflection leaked a number of pixels which still were dominated by the specular RFI outside the model. This can be seen in the image residuals (see fig.\ref{fig: crop_image}) having $\approx 10-15\%$ flux density to the respective quasi-specular models. We used a catalogue of FM transmitters to model the reflected power received at the MWA from all the FM stations at the time of the ON-Moon observation. We made two basic assumptions in order to generate the Earthshine. First, we assumed that the FM stations always remain active and transmit isotropically. Second, we counted all the stations from where the Moon was above the horizon at the time of the observation. Although these assumptions were made due to incomplete information on the beam patterns and operating hours of the FM stations, it certainly limits our analysis of the Earthshine mitigation using simulation (it can be seen in fig.\ref{fig: data_vs_sim_flux} that the observed flux density fluctuates significantly within the assumed bandwidth of the FM stations which otherwise is constant in the simulation). In reality, the FM transmitters beam the signal along the horizontal direction and have a significant signal loss $(\approx 20\rm dB)$  within $10^\circ$ tangential to the beam direction. This directivity and beam pattern of the station can significantly reduce the station counts and allow only stations having a Moon altitude of approximately $10^\circ$ (near the horizon). Taking into account this effect would replace the $4\pi$ factor from the eq.\ref{eq: fm_flux} with the beam angle $\theta$, which can significantly alter the simulated flux density. Using that, we can also calculate the specular reflections using a vector ray tracing algorithm. We can assume a $\approx 16^{\prime\prime}$ region around the centre of the Moon to act as a smooth FM reflector and isolate the FM stations which satisfy the reflection criteria. Therefore, one can argue that the accuracy of the FM simulation would significantly alter based on the assumptions and number of factors contributing to the Earthshine.

\subsection{Earthshine Avoidance}\label{subsection: earthsine_avoidence}
\begin{figure*}[ht]
    \centering
    \includegraphics[scale=0.6]{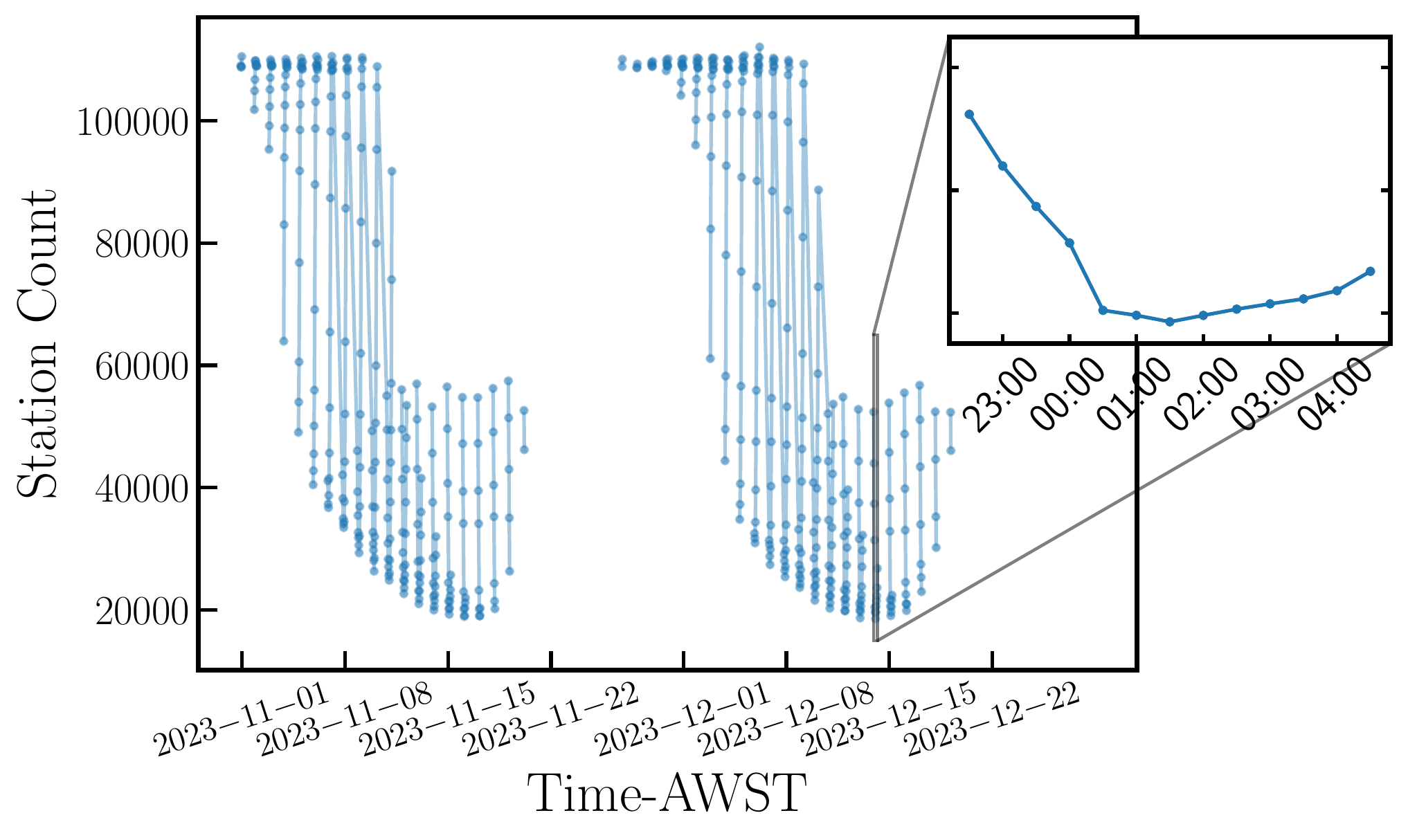}
    \caption{Simulation of the station counts between November and December 2023. The variation of station counts roughly repeats over a month.
    The individual spikes in the plot correspond to a single night's observation between 20:00-4:30 hrs, with a time separation of 30 mins. The zoomed figure on the top right shows the variation of station counts over a single night from 13th December 2023.}
    \label{fig: FM_station_vs_time}
\end{figure*}

As we have seen in our simulation results (see fig. \ref{fig: flux_vs_st_count}), increasing station count increases the overall reflected flux density from the Moon; thus, it can be said that the reflected FM from the Moon correlates with the Earth's terrestrial area (i.e. the land area) being exposed to the Moon. In contrast, a minor reflected FM contribution would be there if the marine parts of the Earth were facing the Moon. Therefore, in Earthshine avoidance, we can use guided simulations to choose suitable observing windows when the FM station count is minimised. This would help us to reduce the strong FM RFI from the Moon, which otherwise hugely contaminates the flux received from the Moon. Fig. \ref{fig: FM_station_vs_time} shows a two-month simulation between November to December 2023 of the FM station count as a proxy for the reflected RFI from the Moon. The simulation targeted only the nighttime between $20:00-04:30$ hrs. as it is relevant for our work and used the location of the Moon as seen by the MWA and FM stations. We counted only those time frames when the Moon was above the horizon at both the MWA and the FM station during the observing window (between $20:00-04:30$ hrs) every night. There are missing data points in the middle of figure \ref{fig: FM_station_vs_time}, which indicate that the Moon is below the horizon at the MWA site during the observing window. The station counts in the figure show repetitive behaviour over a month's time. During certain days, the station counts are significantly less than others; hence, we can utilise such time windows for scheduling future observations of the Moon.

\subsection{Future prospects for 21-cm signal measurements} \label{section: future_prospects}

Our current estimates motivate us to explore other techniques of Earthshine mitigation in our future work. The RFI avoidance technique might be helpful to deal with the strong reflected RFI and, therefore, can provide much-improved estimates of the observable quantities. However, we also wish to investigate whether the inclusion of the station beam patterns improves our modelling and Earthshine mitigation. In our immediate future work, we aim to analyse our latest observation of the Moon using MWA-phase II extended from 2022. MWA-phase II extended has a higher angular resolution, and it can be helpful in improving the total SNR in the FM band. Based on the results, it would be valuable to check whether we can perform foreground mitigation techniques on the phase-II data. We can utilise a similar approach as followed by the SARAS project to quantify the acceptable regimes of the global 21-cm models or alternatively explore Principle Component Analysis (PCA), Gaussian Process Regression (GPR), and Machine Learning (ML) techniques \citep{Tauscher_2021, Makinen_2021, Zuo_2019} to isolate the EoR component from our data.  In addition to that, we can explore new techniques based on the simulations of the FM station counts during the given nights. We can choose two distinct nights where the behaviour of the station counts matches (hence the reflected RFI), train machine learning models with the datasets from one night, and use the trained models to mitigate the Earthshine on the other. However, in such methods, the models are required to be trained up to very high accuracy. It would be worth checking the goodness of such trained models if operated on the datasets from other nights or different instruments (e.g. EDA-2). 

\section{Conclusion} \label{section: conclusion}

The lunar occultation technique aims to utilise radio interferometers to estimate the sky-average global 21-cm EoR signal. However, one must accurately mitigate the Galactic foreground and reflected Earthshine in order to reach the final goal of detecting the global 21-cm signal. In this work, we used two approaches to model the FM Earthshine from the Moon. We started with six nights of LST-locked ON and OFF Moon observations at $40$ kHz fine channel resolution and generated difference images of the Moon between $\approx 70-180$ MHz. We observed that the individual observations around the FM band were highly contaminated by the Earthshine, which otherwise gets suppressed at coarse channels. We assumed two Earthshine components (the diffuse disk and quasi-specular) in the Moon's emission \citep{Ben_2018} and used the relation between the specular and diffuse components from \citet{Evans_1969} to estimate the flux density of the Moon. We took two different approaches to get the final estimates of the flux density of the Moon. In the first approach, similar to \citet{Ben_2018}, we utilised the data itself to mitigate the Earthshine from the data, whereas, in the second approach, we used FM simulations to mitigate the Earthsine from the data. We used an FM catalogue of radio transmitters across the Earth to estimate the simulated diffuse Earthshine from the disk of the Moon. Using these methods, we estimated the flux density of the Moon (fig. \ref{fig: Flux_density_all} bottom panel) and converted it into the brightness temperature (ref. eq. \ref{eq: delta_t}). The brightness temperature is the measure of the difference between the intrinsic temperature of the Moon $(T_{\rm Moon})$ and the Galactic foreground temperature $(T_{\rm Gal}(\nu))$. We have assumed that the Moon had a constant temperature over our desirable frequencies; therefore, we fit the observed brightness temperature from eq.\ref{eq: T_gal-moon_data} with a modified Galactic power-law equation which had an additional factor of constant temperature offset  $(T_{\rm offset})$ (see eq.\ref{eq: T_gal_moon_fit}). The offset temperature measured the $T_{\rm Moon}$. We jointly estimated the $T_{\rm Gal}(\nu)$, $T_{\rm Moon}$ and spectral index $\alpha$ by combining all of the three observation epochs while putting a restriction on the $T_{\rm Moon}$. 
In our joint analysis, we were able to recover the Galactic spectral index $(\alpha)$ of the occulted sky within $5-7\%$ level of the GSM2016 estimates. Also, our estimates of the $T_{\rm Moon}$ ($184.40\pm{2.65} \rm K$ and $173.77\pm{2.48} \rm K$ for the first and second Earthshine mitigation methods, respectively) provided tighter constraints and are consistent with the previous results from \citet{Ben_2018}.

This shows that we can include multiple nights of data and can strengthen the total SNR of our analysis. The next step would be to check the performance of the foreground subtraction techniques on the data. The estimation of the global 21-cm signal is the ultimate goal of our work. Tackling the prior difficulties of these approaches can put us one step closer to understanding the CD-EoR.

\begin{acknowledgement}

This project is supported by an ARC Future Fellowship under grant FT180100321. This research was partially supported by the Australian Research Council Centre of Excellence for All Sky Astrophysics in 3 Dimensions (ASTRO 3D), through project number CE170100013. The International Centre for Radio Astronomy Research (ICRAR) is a Joint Venture of Curtin University and The University of Western Australia, funded by the Western Australian State government.
The MWA Phase II upgrade project was supported by Australian Research Council LIEF grant LE160100031 and the Dunlap Institute for Astronomy and Astrophysics at the University of Toronto.  This scientific work makes use of Inyarrimanha Ilgari Bundara, the CSIRO Murchison Radio-astronomy Observatory. We acknowledge the Wajarri Yamatji people as the traditional owners and native title holders of the Observatory site. Support for the operation of the MWA is provided by the Australian Government (NCRIS), under a contract to Curtin University administered by Astronomy Australia Limited. We acknowledge the Pawsey Supercomputing Centre, which is supported by the Western Australian and Australian Governments. Data were processed at the Pawsey Supercomputing Centre. 

\section{Data Availability}
The data used in this work will be made available upon reasonable request. 
\end{acknowledgement}

\printbibliography

\appendix

\section{Noise Estimation}

\subsection{ For \texorpdfstring{$ S_m(\nu)$}{Lg}} \label{app: S_m_err}

Our analysis used 34 full-band observations (72-230 MHz), with each full-band observation comprising 5 coarse-band observations (in the context of MWA, each coarse-band observation is at 1.28 MHz resolution and 30.72 MHz wide).  We produced the beam-corrected images at every 40~kHz fine channel for each of the observations. As a result, we had $768 \times 5$ beam-corrected images for each full-band observation.  We evaluated the RMS noise for both disk and quasi-specular components $(\Delta S_{\rm disk}, \Delta S_{\rm spec})$ using eq.\ref{eq: tot_flux} and propagated these errors through eq.\ref{eq: moon_flux} to estimate $\Delta S_m(\nu)$. In eq. \ref{eq: moon_flux}, we used the values of $S_{\rm disk}(\nu_0)$ to get $S_{\rm diffuse}(\nu_0)$.  We fitted a line to $S_{\rm disk}(\nu)$ and obtained  $S_m(\nu_0)$ as the fitted value at $\nu_0$. 
\[ S_{\rm diffuse}(\nu_0) = S_{\rm disk}(\nu_0)-Y_{\rm fit}(\nu_0)\]
The errors on the fitting parameters, namely the slope and intercept $(\Delta m, \Delta c)$ of the fitted line, were used to obtain the fitting error of the line at $\nu_0$. 
\[\Delta Y_{{\rm fit}}(\nu_0) = \sqrt{( \Delta m \times \nu_0)^2 + \Delta c^2}\]
therefore, $\Delta S_{\rm diffuse}(\nu_0)$ was evaluated as,
\begin{multline*}
    \Delta S_{\rm diffuse}({\nu_0}) = \\ \sqrt{\Delta S_{\rm disk}^2(\nu_0) + \Delta Y_{\rm fit}^2(\nu_0)}
\end{multline*}
Please note that, as our T-tests satisfy the null hypothesis at $\nu_0$ in all three epochs, we used the same value of uncertainty $\Delta S_{\rm diffuse} (\nu_0)$ in both methods of Earthshine mitigation.  From there, we estimated $\Delta R_e(\nu)$,

\begin{multline}
   \Delta R_e(\nu) =  \left(\frac{\nu}{\nu_0}\right)^{0.58} \times \\ \sqrt{\left(\frac{\Delta S_{\rm diffuse}(\nu_0)}{S_{\rm diffuse}(\nu_0)}\right)^2 + \left(\frac{\Delta S_{\rm spec}(\nu_0)}{S_{\rm spec}(\nu_0)}\right)^2}
\end{multline}

and $\Delta S_{\rm diffuse}(\nu)$ was obtained using,

\begin{multline}
    \Delta S_{\rm diffuse}(\nu) = R_e(\nu)  S_{\rm spec} (\nu) \times \\ \sqrt{\left(\frac{\Delta R_e(\nu)}{R_e(\nu)}\right)^2 + \left(\frac{\Delta S_{\rm spec}(\nu)}{S_{\rm spec}(\nu)}\right)^2} 
\end{multline}

Finally, $\Delta S_m(\nu)$ was estimated as,
\begin{equation}
    \Delta S_m(\nu) = \sqrt{\Delta S_{\rm disk}^2(\nu) + \Delta S_{\rm diffuse}^2 (\nu)}
\end{equation}
As a final result, we had $S_m(\nu)$ and  $\Delta S_m(\nu)$ for all 34 full-band observations and propagated them as quadrature in the mean. We put the estimated errors in eq.\ref{eq: delta_t} to estimate the occulted sky temperature and errors therein.

\subsection{ For GSM \texorpdfstring{$T_{\rm Gal}(\nu)$}{Lg} and \texorpdfstring{$T_{\rm refl-Gal}(\nu)$}{Lg} models}\label{app: Tgal_Tref_gal_err}

We used GSM models to generate the sky maps in the middle of each 34 full-band observations. The maps were generated to match the observed frequency range with a frequency resolution of 5 MHz.
In order to get the Galactic temperature $(T_{\rm Gal150})$ and Galactic spectral index $(\alpha)$, we fit our model with a power-law equation
\begin{equation}
    \label{eq: T_gal}
    T_{\rm Gal}(\nu) = T_{\rm Gal150}\left(\frac{\nu}{150\rm MHz}\right)^\alpha
\end{equation}
We used the GSM2016 model to represent the Galactic temperature. We used the model's $5\%$ intrinsic map estimation error in the fitting and measured the values of $T_{\rm Gal150}$ and $\alpha$ for all 34 full-band observations. The fitted parameters $(T_{\rm Gal150}, \alpha)$ and the uncertainties on the fitted parameter $(\Delta T_{\rm Gal150}, \Delta \alpha)$ values were propagated in the quadrature rule. In a similar fashion, we estimated the uncertainties in the $T_{\rm refl-Gal}(\nu)$ using the eq. \ref{eq: refl_temp} (to distinguish between the spectral index parameter of the reflected Galactic power-law equation and the Galactic power-law equation, we denoted the reflected spectral index parameter in eq. \ref{eq: refl_temp} as $\beta$). 

\subsection{Reflected flux density variation with station count} \label{app: ref_flux_station}

\begin{figure}[ht]
    \centering
    \includegraphics[scale=0.38]{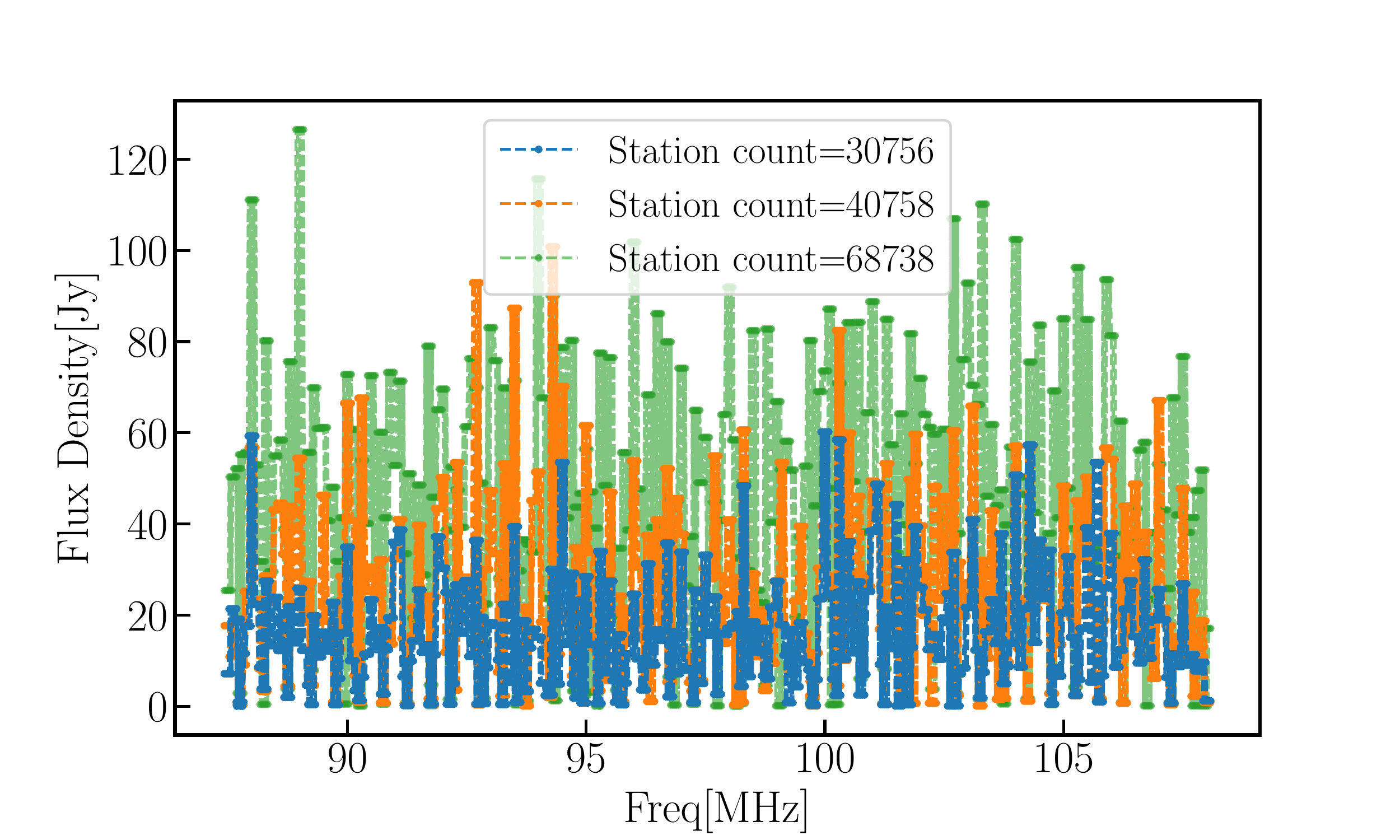}
    \caption{Reflected FM flux density at three different time stamps during the second observing epoch.}
    \label{fig: flux_vs_st_count}
\end{figure}

We obtained the reflected FM flux density corresponding to every ON-Moon observation comprising the FM band. The variation of the reflected FM flux density changes with the number of FM stations. The increasing number of FM stations can be understood as the more terrestrial land area being exposed to the Moon. Fig. \ref{fig: flux_vs_st_count} showing the variation in the reflected flux density at three-time stamps during the second observing epoch.

\subsection{T-test results} \label{app: p_value_analysis}
\begin{figure}[ht]
    \centering
    \includegraphics[scale=0.40]{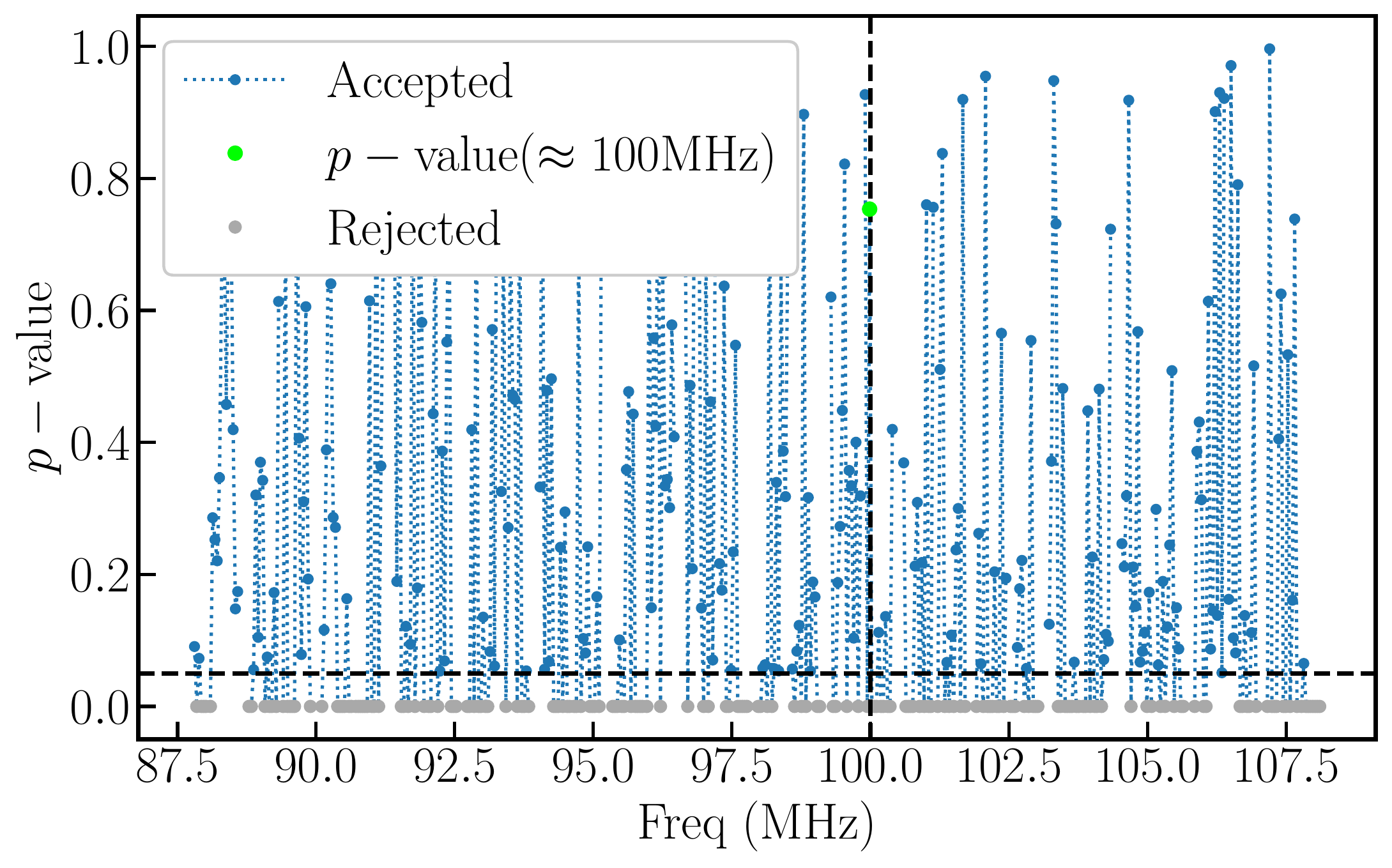}
    \caption{T-test results on the first epoch datasets. X-axis shows the FM frequency band, and the Y-axis shows the corresponding $p$-values. 
    The horizontal line is at $p=0.05$, above which the null hypothesis is considered to be \textit{accepted}. The data points shown in \textit{blue} colour show the \textit{accepted}, and the \textit{grey} colour show \textit{rejected} frequency channels. The $p-$value at $\approx \nu_0$ is shown with \textit{lime} colour. }
    \label{fig: p_value_plot}
\end{figure}

We estimated the $p$-values from the T-test at all fine-frequency channels . As we already mentioned the limitation of our simulations, our T-tests performed poorly on more than half of the fine-frequency channels. The null hypothesis got rejected at those channels. However, we only required a single value of $S_{\rm diffuse}$ at $\nu_0$ to estimate $R_e(\nu)$ and since our T-test accepted the null hypothesis at $\approx \nu_0$ at all three epochs, we utilised the values of simulated reflected flux density $S_{\rm FM}(\nu_0)$ as $S_{\rm diffuse}(\nu_0)$. Fig. \ref{fig: p_value_plot} showing the T-test results on the dataset from the first epoch.

\subsection{Correlation between the fitted parameters} \label{app: correlation}
\begin{table}
    \centering
    \begin{tabular}{c|c|c|c}
         &  $T_{\rm Gal150}$ & $\alpha$ & $T_{\rm offset}$\\
         \hline
      $T_{\rm Gal150}$   & 1 & 0.99 & 0.99 \\
      $\alpha$ & . &1 & 0.99 \\
      $T_{\rm offset}$ & . & . & 1 \\
    \end{tabular}
    \caption{Correlation between the parameters of eq. \ref{eq: T_gal_moon_fit} when fitted with the dataset from the first epoch using the first Earthshine mitigation method. }
    \label{tab: corr_table_epoch_one1}
\end{table}
\begin{table*}[ht]
    \centering
    \begin{tabular}{c|c|c|c|c|c|c|c}
        & $T_{\rm Gal150;~Aug.}$ & $T_{\rm Gal150;~Sept.}$ & $T_{\rm Gal150;~Dec.}$ & $\alpha_{\rm fitting;~Aug.}$ & $\alpha_{\rm fitting;~Sept.}$ & $\alpha_{\rm fitting;~Dec.}$ & $T_{\rm offset}$  \\
        \hline
        $T_{\rm Gal150;~Aug.}$ &1.000&0.965&0.954&0.914&0.797&0.806&0.992  \\
        $T_{\rm Gal150;~Sept.}$ &.&1.000&0.932&0.911&0.819&0.790&0.973 \\
        $T_{\rm Gal150;~Dec.}$ &.&.&1.000&0.897&0.770&0.848&0.958\\
        $\alpha_{\rm fitting;~Aug.}$ &.&.&.&1.000&0.753&0.761&0.936\\
        $\alpha_{\rm fitting;~Sept.}$ &.&.&.&.&1.000&0.653&0.804\\
        $\alpha_{\rm fitting;~Dec.}$ &.&.&.&.&.&1.000&0.812\\
        $T_{\rm offset}$ &.&.&.&.&.&.&1.000\\
    \end{tabular}
    \caption{The correlation between the parameter of eq. \ref{eq: T_gal_moon_fit} when fitted jointly with combined datasets from all three epochs.}
    \label{tab: joint_corr_table}
\end{table*}

Our final estimates of occulted sky temperature $T_{\rm Gal150}$, Galactic spectral index $\alpha$, and offset temperature $T_{\rm offset}$ (which represents the $T_{\rm Moon}$) show a very high correlation with each other. 
When checked on the single night dataset (from fig. \ref{fig: gal_temp_with_FM_rem_epoch1}), it showed a near unity level of correlation, see table \ref{tab: corr_table_epoch_one1}.

The sky position of the Moon is different at all three observing epochs; hence, the galactic spectral index and occulted sky temperature are also different at all three epochs. In the joint analysis, we considered these parameters as independent parameters while restricting $T_{\rm offset}$ (assuming the Moon to have the same temperature across the epochs). The correlation table (\ref{tab: joint_corr_table}) between all parameters still shows a high correlation, however, the spectral index shows less correlation compared to the temperature parameters.

\begin{table*}
    \centering
    \begin{tabular}{c|c|c|c|c|c|c|c}
    \toprule
    
     $\rm Epoch$  & $\rm Sky-model$ & $T_{\{\rm Gal150;~ model\}}\rm ~(K)$ & $\alpha_{\rm model}$ & $\rm Fitting$ & $T_{\{\rm Gal150;~fitting\}}\rm ~(K)$ &  $\alpha_{\rm fitting}$  &  $T_{\{\rm Moon\}}\rm ~(K)$  \\
        \midrule
        & $\rm GSM$ &  $250.4\pm{12.5}$ & $-2.540 \pm{0.002}$ & $\rm Method~ 1$ & $199.2\pm{5.4}$   &   $-2.70 \pm{0.05}$ & $188.3 \pm{4.6}$\\
        &  &  &  & $\rm Method~ 2^{\rm ~(FM~sim.)}$& $183.5\pm{5.1}$ &   $-2.79 \pm{0.05}$ & $174.2\pm{4.4}$\\
        
          & $\rm GSM2016$ & $242.7\pm{12.1}$& $-2.621\pm{0.003}$& . & $195.5\pm{5.3}$ & $-2.72\pm{0.05}$ & $186.1\pm{4.5}$ \\
         $\rm Aug.$ & & & & ..&$179.2\pm{5.0}$ & $-2.82\pm{0.05}$ & $171.4\pm{4.3}$ \\
 
 & $\rm LFSM$ & $313.9\pm{15.7}$ & $-2.689\pm{0.003}$&.& $202.8\pm{5.3}$  &  $-2.70\pm{0.05}$  & $188.5\pm{4.5}$ \\
           &  &  & & ..& $189.0\pm{5.1}$    & $-2.78\pm{0.05}$     & $176.2\pm{4.4}$ \\
          
          & $\rm Haslam$ &  $253.7\pm{12.7}$ &  $-2.603\pm{0.003}$ & . &$199.2 \pm{5.3}$  &  $-2.71 \pm{0.05}$ & $187.9 \pm{4.6}$\\
        & &   & &.. &  $183.7 \pm{5.1}$    &  $-2.80 \pm{0.05}$ & $174.0 \pm{4.3}$\\

          \midrule
          
       & $\rm GSM$ &  $253.4\pm{12.7}$   & $-2.540 \pm{0.002}$  &. & $177.6\pm{4.4}$&   $-2.54 \pm{0.04}$ & $189.4 \pm{4.2}$\\
        &  &  && ..&   $185.6\pm{5.1}$   &  $-2.43 \pm{0.05}$ & $198.5\pm{4.9}$\\
        
          & $\rm GSM2016$ & $241.0\pm{12.0}$ & $-2.585\pm{0.003}$ &. &$174.4\pm{4.3}$ & $-2.56\pm{0.04}$ & $187.4\pm{4.4}$ \\
         $\rm Sept.$ & & & &..& $182.2\pm{5.0}$ &  $-2.47\pm{0.05}$ & $196.0\pm{4.8}$ \\
 
 & $\rm LFSM$ & $295.3\pm{14.8}$  & $-2.689\pm{0.003}$  &.& $180.8\pm{4.4}$ & $-2.56\pm{0.04}$  & $189.2\pm{4.2}$ \\
           &  &  & & ..&$188.6\pm{5.0}$ &    $-2.46\pm{0.04}$     & $198.0\pm{4.8}$ \\
          
          & $\rm Haslam$ &  $250.7\pm{12.5}$  &  $-2.603\pm{0.003}$ &.&  $177.4 \pm{4.4}$ &  $-2.56 \pm{0.04}$ & $188.9 \pm{4.2}$\\
        & &   & &.. & $185.2 \pm{5.0}$ &    $-2.45 \pm{0.05}$  &$197.8 \pm{4.8}$\\
        
           \midrule
        & $\rm GSM$ &  $395.8\pm{19.8}$   & $-2.496 \pm{0.002}$  & .& $235.6\pm{5.4}$ &   $-2.66 \pm{0.04}$ & $173.8 \pm{5.0}$\\
        &  &  & &..&   $239.5\pm{4.1}$ &  $-2.60 \pm{0.03}$ &$177.4\pm{3.8}$\\
        
          & $\rm GSM2016$ & $380.4\pm{19.0}$ & $-2.497\pm{0.003}$&.&  $232.1\pm{5.4}$ &$-2.68\pm{0.04}$ & $172.2\pm{5.0}$ \\
         $\rm Dec.$ & & & &..& $235.7\pm{4.0}$  & $-2.62\pm{0.03}$ & $175.5\pm{3.7}$ \\
 
 & $\rm LFSM$ & $453.4\pm{22.7}$  & $-2.557\pm{0.003}$ &.& $238.7\pm{5.4}$ & $-2.68\pm{0.04}$  & $173.7\pm{5.0}$ \\
           &  &  & &..& $242.4\pm{4.0}$ &    $-2.62\pm{0.03}$    & $177.2\pm{3.7}$ \\
          
          & $\rm Haslam$ &  $414.2\pm{20.6}$  &  $-2.603\pm{0.003}$ &. & $236.0 \pm{5.4}$ &  $-2.68 \pm{0.04}$ & $173.4 \pm{5.0}$\\
        & &   & &..& $239.7 \pm{4.0}$ &    $-2.62 \pm{0.03}$  &$176.8 \pm{3.7}$\\
        
    \bottomrule
    \end{tabular}
    \caption{Table showing the best-fit parameters from all sky models for every individual case.}
    \label{tab: table_individual_all_models}
\end{table*}

\begin{table*}

    \centering
    
    \begin{tabular}{c|c|c}
    \toprule
    
     $\rm Sky-model$ & $T_{\{\rm Moon\}}\rm ~(K)$ & $\rm Fitting$ \\
        \midrule
        $\rm GSM$ & $186.5 \pm{2.7}$ & $\rm Method~ 1$\\ 
        &  $176.7\pm{2.5}$ & $\rm Method~ 2^{\rm ~(FM~sim.)}$\\ 
        
          $\rm GSM2016$  & $184.4\pm{2.6}$& .\\
          &  $173.8\pm{2.5}$& ..\\
 
  $\rm LFSM$ & $186.6\pm{2.7}$& .\\
           & $178.7\pm{2.5}$& ..\\
          
           $\rm Haslam$ & $186.1 \pm{2.7}$& .\\
        &  $176.4 \pm{2.5}$& ..\\
    \bottomrule
    \end{tabular}
    \begin{tabular}{c|c|c|c}
    \toprule
    
     $\rm Epoch$ & $T_{\{\rm Gal150\}}\rm ~(K)$& $T_{\{\rm Moon\}}\rm ~(K)$ & $\rm Joint ~Fitting$ \\
        \midrule
        $\rm Aug.$ & $196.9 \pm{6.1}$ & $187.6 \pm{5.3}$ &$\rm Method~ 1$\\ 
        &  $179.5\pm{5.6}$ &$171.3 \pm{4.8}$ &  $\rm Method~ 2^{\rm ~(FM~sim.)}$\\ 
        
           $\rm Sept.$ & $174.8 \pm{4.8}$ & $187.7 \pm{4.6}$ & .\\ 
        &  $181.3\pm{4.9}$ &$195.1 \pm{4.7}$ &   ..\\ 
        
  $\rm Dec.$ & $223.3 \pm{5.3}$ & $164.5 \pm{4.9}$ &.\\ 
        &  $236.2\pm{4.3}$ &$175.9 \pm{4.0}$ & ..\\ 
\hline
          $\rm Joint~ Aug.$ & $192.6 \pm{3.4}$ & $184.5 \pm{2.9}$ &$\rm Method~ 1$\\ 
          $\rm Sept.$ & $171.5 \pm{3.2}$ & '' &.\\ 
          $\rm Dec.$ & $243.2 \pm{3.3}$ &'' &.\\ 
          \hline
        
        $\rm Joint~ Aug.$&  $178.8\pm{3.0}$ &$173.3 \pm{2.6}$ &  $\rm Method~ 2^{\rm ~(FM~sim.)}$\\ 
        $\rm Sept.$&  $159.1\pm{2.8}$ &'' &  ..\\ 
       $\rm Dec.$ &  $231.5\pm{2.9}$ &'' &  ..\\ 
        
    \bottomrule
    \end{tabular}
    \caption{Left: Joint-fit estimates of $T_{\rm Moon}$ from all sky models. Right: GSM2016 model fit estimates for removed FM-band.}
    \label{tab: table_joint_all_models}
\end{table*}

\end{document}